\documentclass[a4paper,11pt]{article}
\pdfoutput=1 % if your are submitting a pdflatex (i.e. if you have
\usepackage{jheppub} % for details on the use of the package, please
\usepackage[toc,page]{appendix}

\usepackage[T1]{fontenc} % if needed
\usepackage{cancel}
\usepackage{soul}

\newcommand{\unit}{\leavevmode\hbox{\small1\kern-3.6pt\normalsize1}}

\definecolor{marron}{rgb}{0.59, 0.29, 0.0}

\newcommand{\gsim}{\lower.7ex\hbox{$\;\stackrel{\textstyle>}{\sim}\;$}}
\newcommand{\lsim}{\lower.7ex\hbox{$\;\stackrel{\textstyle<}{\sim}\;$}}

\newcommand{\be}{\begin{equation}}
\newcommand{\ee}{\end{equation}}

\newcommand{\bea}{\begin{eqnarray}}
\newcommand{\eea}{\end{eqnarray}}

\def\simlt{\stackrel{<}{{}_\sim}}

\title{Anomaly-free Dark Matter with Harmless Direct Detection Constraints}

\author[a,b]{S.~Caron }
\author[c]{J.~A.~Casas }
\author[c]{J. ~Quilis}
\author[d]{R.~Ruiz de Austri}

\affiliation[a]{Institute for Mathematics, Astrophysics and Particle Physics, Faculty of Science, Mailbox 79, Radboud University Nijmegen, P.O. Box 9010, NL-6500 GL Nijmegen, The Netherlands}
\affiliation[b]{Nikhef, Science Park, Amsterdam, The Netherlands}
\affiliation[c]{Instituto de F\'{\i}sica Te\'{o}rica, IFT-UAM/CSIC, Universidad Aut\'{o}noma de Madrid,  28049, Madrid, Spain}
\affiliation[d]{Instituto de F\'{i}sica Corpuscular, IFIC-UV/CSIC, Valencia, Spain}

\emailAdd{scaron@cern.ch}
\emailAdd{j.alberto.casas@gmail.com}
\emailAdd{javier.quilis@csic.es}
\emailAdd{rruiz@ific.uv.es}

\keywords{Beyond Standard Model Phenomenology, Dark Matter, LHC}

\abstract{
Dark matter (DM) interacting with the SM fields via a $Z'-$boson (`$Z'-$portal') remains one of the most attractive WIMP scenarios, both from the theoretical and the phenomenological points of view. In order to avoid the strong constraints from direct detection and dilepton production, it is highly convenient that the $Z'$ has axial coupling to DM and leptophobic couplings to the SM particles, respectively. The latter implies that the associated $U(1)$  coincides with baryon number in the SM sector.
In this paper we  completely classify the possible anomaly-free leptophobic $Z'$ with minimal dark sector, including the cases
where the coupling to DM is axial. The resulting scenario is very predictive and perfectly viable from the present constraints from DM detection, EW observables and LHC data (di-lepton, di-jet and mono-jet production). We analyze all these constraints, obtaining the allowed areas in the parameter space, which generically prefer $m_{Z'}\simlt 500$ GeV, apart from resonant regions. The best chances to test these viable areas come from future LHC measurements. 
}

\begin{document} 
%%%%%%%%%%%%%%%%%%%%%%%%%%%%%%%

\rightline{IFT-UAM/CSIC-18-82}

\maketitle
\flushbottom
%%%%%%%%%%%%%%%%%%%%%%%%
\section{Introduction}
\label{sec:introduction}
%%%%%%%%%%%%%%%%%%%%%%%%%

The simplest WIMP models for dark matter (DM), where the dark sector consists of one single particle interacting with the SM fields via Higgs$-$ or $Z-$boson (i.e. Higgs and $Z$ portals) are currently under pressure, especially by DM direct detection (DD) experiments.

However, this view is probably over-simplified, in several ways. First,  the dark sector may consists of several particles (even if only one of them is the DM). Second, the dark sector may not be directly coupled to the SM one, but through some mediator, e.g. a new scalar or a new vector boson, $Z'$. Models of the last kind have been extensively considered in the  literature \cite{Langacker:1984dc, Langacker:2008yv, FileviezPerez:2010gw, Frandsen:2011cg, Duerr:2013dza, Duerr:2013lka, Alves:2013tqa, Arcadi:2013qia, Lebedev:2014bba, Duerr:2014wra, Perez:2015rza, Duerr:2015vna, Kahlhoefer:2015bea, Jacques:2016dqz, Fairbairn:2016iuf, Arcadi:2017hfi, Perez:2014qfa, Ohmer:2015lxa, Ismail:2016tod, Okada:2017dqs, Okada:2016tci, Bandyopadhyay:2018cwu}, as they represent a very plausible scenario of BSM physics, e.g. in the context of GUT or string models. Usually, the analyses have been done in the framework of  the so called simplified DM models (SDMM), where  the DM particle and the $Z'$ mediator are the only extra fields. Still, there is a non-trivial parameter space, essentially given by the $Z'-$mass, its coupling to the DM particle, and the various couplings to the SM fields. Some of the most important constraints in that parameter-space come from DD experiments \cite{Duerr:2014wra, Arcadi:2017hfi} and from di-lepton production at the LHC \cite{Kahlhoefer:2015bea, Arcadi:2017hfi}. These constraints are highly alleviated if the coupling of the $Z'$ with DM is of the axial type, and if the $Z'$ has leptophobic couplings to the SM particles, respectively.

On the other hand, as stressed in  several articles \cite{FileviezPerez:2010gw, Duerr:2013dza,Duerr:2013lka,Duerr:2014wra,Perez:2014qfa, Perez:2015rza,Duerr:2015vna,Kahlhoefer:2015bea, Ellis:2017tkh},   simplified DM models are ``too simple" concerning unitarity, gauge invariance and anomaly cancellation. 
In fact, the $Z$'s in SDMM are typically anomalous.
%
%{\red SC: here one needs to define it a bit more what is meant. Ref. 7 speaks mainly about unitarity problems, whereas ref 12 is about triangular anomalies and cancellation of the hypercharges. AC: There was a mistake in the ref. number. I have added more explanation (in blue)}
Then, in order to cancel the anomalies, additional fermions
%{\red (SC: i.e. fermions for the triangular anomalies) AC: corrected}
(besides the DM one) are mandatory. The authors  of ref. \cite{FileviezPerez:2010gw, Duerr:2013dza,Duerr:2013lka,Duerr:2014wra,Perez:2014qfa, Perez:2015rza,Duerr:2015vna, Ellis:2017tkh}  performed a systematic search of (anomaly-free) $Z'$ extensions either with axial DM-coupling or with leptophobia  (or, equivalently, completions of gauged baryon number).
%(though they do not single-out models accomplishing both conditions simultaneously).
%{\red SC: (Is that actually possible? It is not for axial-axial couplings according to Ref 7 and requires SU2W according to Ellis et al paper. We need to comment on this here.) AC: It is possible because the coupling of $Z'$ to SM is vectorial (it is like baryon number). This is explained in sect.2.}
In this paper we follow a similar spirit, obtaining new general results on this type of consistent $Z'$ extensions. We will assume throughout the paper that the DM particle is a Dirac fermion, $\chi$, neutral under all the SM gauge symmetries. Then, we will determine the possible scenarios where the $Z'$ is simultaneously leptophobic and with %{\red SC: (maybe change to:) has a pure $U(1)_{Y'}$ axial coupling, which would be the case if the $\chi$ would be a Majorana fermion. AC: the coupling is not pure axial in the sense that is axial to DM but vectorial to SM fields.} 
axial DM coupling. There are very few scenarios of that kind with a minimal dark spectrum. Finally, we study the phenomenology of these models, and discuss how they can be experimentally tested. 

Our paper is structured as follows. In Secs. \ref{sec:model} and \ref{sec:axialmodel} we present the model. In Sec. \ref{sec:pheno} we illustrate the relevant constraints that apply to our model from electroweak precision measurements, LHC, DM relic density and direct and indirect DM searches. In Sec. \ref{sec:results} we illustrate our results, and in Sec. \ref{sec:conclusions} we summarize our conclusions.

%%%%%%%%%%%%%%%%%%%%%%%%%%%%%%%%%%%%%%%%%%
\section{Anomaly-free leptophobic $Z'$s}
\label{sec:model}
%%%%%%%%%%%%%%%%%%%%%%%%%%%%%%%%%%%%%%%%%%

It is easy to see that a consistent leptophobic $U(1)_{Y'}$ group, where leptons have vanishing $Y'-$charge, must be equivalent to baryonic number, $U(1)_B$, in the SM sector. The invariance of the leptonic Yukawa couplings,
\be
y^e_i \bar L_i H e_i ,
\ee
(where $y_i$ are the Yukawa coupling constants, with $i$ a family index in an obvious notation) requires the $Y'-$charge of the Higgs to vanish, $Y'_H=0$.
%, as we start with $Y'_{L_i}=Y'_{e_i}=0$. 
Then, invariance of the hadronic Yukawa couplings
\be
y^u_{ij} \bar Q_i\bar H u_j , \ \ \ \  y^d_{ij} \bar Q_i H d_j
\ee
requires $Y'_{Q}=Y'_{u}=Y'_{d}$, which is equivalent to $U(1)_B$.
%unless different families have different $Y'$ charges, which is extremely problematic from FCNC. 
So, in the following we will assume $U(1)_{Y'}\equiv U(1)_B$ in the SM sector, and therefore $Y'=1/3$ for all quarks. Note that this is a completely generic result for any UV completion of the SM with a leptophobic, flavour-blind, $U(1)_{Y'}$ group. 

A consequence of the previous result is that a (leptophobic) $Z'$ couples to quarks in a purely vectorial way. This has important implications, especially for DD experiments. Namely, if the $Z'$ couples also in a vectorial way to DM, then the effective operator for DD is spin-independent with no velocity-supression. Hence the model would be under extreme pressure from DD bounds as it has been shown for instance in Ref. \cite{Kahlhoefer:2015bea, Arcadi:2017hfi}. On the other hand, if the $Z'$ coupling to DM is axial, then the effective DD operator is both spin-dependent and velocity-suppressed, so the model is safe from DD bounds. We will come back to this point in Sec. \ref{sec:pheno}. Next, we examine further conditions imposed by the requirement of leptophobia.

Since  $U(1)_{Y'}\equiv U(1)_B$ for the SM fields, there are  two anomalies\footnote Previous systematic studies on anomaly cancellation conditions for $U(1)_B$ extensions have been performed in refs. \cite{Pais:1973mi,Rajpoot:1987yg,Foot:1989ts,Carone:1995pu,Georgi:1996ei,Dulaney:2010dj,FileviezPerez:2010gw,FileviezPerez:2011pt,Arnold:2013qja,Duerr:2013lka,Perez:2013tea,Batell:2014yra,Duerr:2014wra,Perez:2015rza,Duerr:2015vna}). which are not vanishing just within the SM sector, and thus require extra staff: $SU(2)_L^2 \times U(1)_{Y'}$ and $U(1)_Y^2 \times U(1)_{Y'}$. The first one requires the presence of non-trivial representations under $SU(2)_L$. Since by assumption, the DM particle, $\chi$, is a SM singlet, the most economical extension is to add two $SU(2)_L$ doublets, $\psi_L, \psi_R$ (the need of at least two of such doublets is obliged e.g. by the cancellation of Witten's $SU(2)$ global anomaly). The cancellation of the anomaly requires
\be
SU(2)_L^2 \times U(1)_{Y'}\ \ {\rm  anomaly} \ \ \longrightarrow \ \ 
Y'_{\psi_L}-Y'_{\psi_R}=-3 .
\label{SU2Yp1}
\ee
Then, it is straightforward to check that the cancellation of the $U(1)_Y^2 \times U(1)_{Y'}$ anomaly 
demands extra particles. Otherwise, such cancellation would require\footnote{We use a normalization of the hypercharge, so that it coincides with the electric charge for $SU(2)_L-$singlets.} $Y_{\psi_L}^2Y'_{\psi_L}-Y_{\psi_R}^2Y'_{\psi_R}=3/4$. 
In addition, the vanishing of the $U(1)_{Y}^3$ anomaly would impose $Y_{\psi_L}=Y_{\psi_R}$. These two conditions, together with Eq. (\ref{SU2Yp1}), lead to $Y_{\psi_L}^2=-1/4$, with no solution. In consequence, we need to add at least one extra singlet, $\eta$, to the dark sector. 
In other words, the minimal dark sector for a leptophobic $Z'$ is:
\be
{\rm  minimal\ dark\ sector:} \ \ \ \{\chi_{L,R},\  \psi_{L,R},\  \eta_{L,R}\} ,
\label{minDS}
\ee
where $\chi$ is a SM singlet (and the DM particle), $\psi$ is a $SU(2)_L$ doublet (and color singlet), and $\eta$ is $SU(2)_L$ and color singlet.

Next, we re-examine the conditions imposed on the charges of the dark sector by the cancellation of the various anomalies:
\be
SU(2)_L^2 \times U(1)_{Y'}\ \ {\rm  anomaly} \ \ \longrightarrow \ \ 
Y'_{\psi_R}=3+Y'_{\psi_L} ,
\label{SU2Yp}
\ee
\be
SU(2)_L^2 \times U(1)_{Y}\ \ {\rm  anomaly} \ \ \longrightarrow \ \ 
Y_{\psi_L}=Y_{\psi_R} \equiv Y_\psi ,
\label{SU2Y}
\ee
\be
U(1)_Y^3 \ {\rm and} \ U(1)_{Y}\ \ {\rm  anomalies} \ \ \longrightarrow \ \ 
Y_{\eta_L}=Y_{\eta_R} \equiv Y_\eta ,
\label{Y3}
\ee
\be
U(1)_Y^2 \times U(1)_{Y'}\ \ {\rm  anomaly} \ \ \longrightarrow \ \ 
Y_{\eta}^2 (Y'_{\eta_L}-Y'_{\eta_R})=\frac{3}{2}+6Y_{\psi}^2  ,
\label{Y2Yp}
\ee
\be
U(1)_{Y'}^2 \times U(1)_{Y}\ \ {\rm  anomaly} \ \ \longrightarrow \ \ 
2Y_{\psi} ({Y'_{\psi_L}}^2-{Y'_{\psi_R}}^2)=
 -Y_{\eta} ({Y'_{\eta_L}}^2-{Y'_{\eta_R}}^2) ,
\label{Yp2Y}
\ee
\be
U(1)_{Y'}\ \ {\rm  anomaly} \ \ \longrightarrow \ \ 
({Y'_{\chi_L}} + {Y'_{\eta_L}}) - ({Y'_{\chi_R}} + {Y'_{\eta_R}})=6 ,
\label{Yp}
\ee
\be
U(1)_{Y'}^3\ \ {\rm  anomaly} \ \ \longrightarrow \ \ 
({Y'_{\chi_L}}^3 + {Y'_{\eta_L}}^3+2 {Y'_{\psi_L}}^3) - ({Y'_{\chi_R}}^3 + {Y'_{\eta_R}}^3+2 {Y'_{\psi_R}}^3)=0 .
\label{Yp3}
\ee

\vspace{0.2cm}
\noindent
Eqs. (\ref{SU2Yp}-\ref{Yp}) can be solved analytically in a straightforward way, leaving $\{ Y_{\psi}, Y_\eta, Y'_{\psi_R}, Y'_{\chi_R}\}$ as the remaining unknowns. Furthermore, $Y_{\psi}, Y_\eta$ are chosen so that the corresponding electric charges are integer, to avoid cosmological disasters. This requires them to be  $m+1/2$ and $n$ respectively, with $m,n$ integers. Then for each choice of $\{ Y_{\psi}, Y_\eta\}$, there is a continuum of consistent values of $\{ Y'_{\psi_R}, Y'_{\chi_R}\}$, although only two (or one in some cases) out of them present axial coupling of the $Z'$ to the DM particle, $\chi$, i.e. $Y'_{\chi_L}=-Y'_{\chi_R}$ (for details and explicit expressions see Appendix A). Besides, only for four special  choices of $\{ Y_{\psi}, Y_\eta\}$, the axial solutions correspond to rational $Y'-$charges (which actually happen to be identical in the four cases), namely
\bea
\label{leptophobax}
\left\{  Y_{\psi} , Y_\eta\right\} &=& \left\{  \pm \frac{1}{2}, \pm 1\right\} ,\  \left\{  \pm \frac{7}{2}, \pm 5\right\} ,
\nonumber
\\
\left\{   Y'_{\psi_L}, Y'_{\psi_R}, Y'_{\eta_L}, Y'_{\eta_R},Y'_{\chi_L}, Y'_{\chi_R}\right\} &=& \left\{-\frac{3}{2}, \frac{3}{2}, \frac{3}{2}, -\frac{3}{2}, \frac{3}{2}, -\frac{3}{2} \right\} . 
\eea

In addition, recall that all quarks have $Y'=1/3$, i.e. their baryon number.
%
%We scan the latter imposing some conditions to avoid preposterous numbers: $Y'_{\psi_L}, Y'_{\chi_L}$ are taken as $a/b$ and $c/d$ respectively, with $a, b, c, d$ integer numbers between $-10$ and $10$. Furthermore, $Y_{\psi}, Y_\eta$ are chosen so that the corresponding electric charges are integer, to avoid cosmological disasters. This requires them to be {\blue $m+1/2$ and $n$} respectively, with $m,n$ integers. In addition we demand the electric charge to be $|Q_{\rm el}|\leq 5$. Then, for each choice of these parameters, Eq. (\ref{Yp3}) is probed. 
%
%In this way we obtain 458 leptophobic solutions to Eqs. (\ref{SU2Yp}-\ref{Yp3}) with minimal dark sector. This includes a subset of four solutions for which the coupling of the $Z'$ to the DM particle, $\chi$, is of axial type: 
%

%%%%%%%%%%%%%%%%%%%%%%%%%%%%%%%%%%%%%%%%%%%%%%%%%%%%%%%%%%%%%%%%%%%%
\section{Anomaly-free leptophobic  $Z'$, with axial coupling to DM}
\label{sec:axialmodel}
%%%%%%%%%%%%%%%%%%%%%%%%%%%%%%%%%%%%%%%%%%%%%%%%%%%%%%%%%%%%%%%%%%%

As mentioned in the previous section, the requirement of axial coupling of the $Z'$ mediator to DM has been advocated  to diminish the pressure of DD bounds on the viability of the scenario. For example, in Ref. \cite{Lebedev:2014bba}, a $Z'$ with axial couplings to both the SM fields and the DM particle, was considered. In this way the $Z'-$mediation leads to spin-dependent effective operators for DD, which are much less constrained. However,  as we have seen, if the $Z'$ is leptophobic (which is desirable), then the coupling to the SM fields is vectorial, since $U(1)_{Y'}$ is equivalent to baryonic number in the observable sector. Hence a leptophobic $Z'$ with axial DM coupling leads to effective operators 
\be
\bar q \gamma_\mu q\ \bar \chi \gamma_5\gamma^\mu \chi ,
\ee
where $q$ is a generic quark. Such operators induce DD interactions that are not only spin-dependent, but also velocity-suppressed. Consequently DD virtually does not impose constraints on a generic leptophobic $Z'$, axially coupled to DM. These are of course good news for this kind of scenario.

An interesting fact is that, assuming minimal DM sector, a leptophobic, DM-axial $Z'$ has completely determined $Y'$ charges for both SM and dark fields, as shown in Eq.  (\ref{leptophobax}).  This means that a usual parameter in SDMM, namely the relative strength of the SM and the DM $Z'-$couplings, is not free anymore. Consequently, a future detection of the $Z'$ mediator at the LHC would also test this scenario. To be more precise, the absolute value of the charge of the DM particle, $\chi$, is 4.5 times larger than that of quarks. Actually, this goes in the right direction to explain why such $Z'$ has not been discovered yet (if it exists, of course): the smaller the couplings to the quarks, the more suppressed the $Z'$ production at the LHC.

Another relevant point has to do with baryon number violation. Since the SM baryonic number is being promoted to an anomaly-free gauge symmetry, which is spontaneously broken (so that the $Z'$ is massive), one should be concerned by baryon-number-violation constraints. The most important of those are proton decay and neutron-antineutron oscillations. Proton decay cannot take place in this context since it needs lepton-number violation as well. On the other hand, neutron-antineutron oscillations represent a violation of baryon number in two units. However, from Eq.(\ref{leptophobax}), it is clear that the scalar field breaking $U(1)_{Y'}$, say $S$, must have $Y'_S=\pm 3$, in order to trigger masses for the dark fields. Consequently, it is not possible to build an effective operator able to mediate neutron-antineutron oscillations. Incidentally, this argument also applies to proton decay, which needs $\Delta B = -1$. 
%It should be noticed that the absence of dangerous baryon-number-violation processes is very generic for leptophobic models (even if they do not have axial couplings), since, from Eq. (\ref{SU2Yp}), $Y'_S=\pm 3$ is obliged to give mass to the $\psi-$doublets.

In order to explore further the phenomenology of leptophobic, DM-axial, $Z'$s, we will focus on one of the four models of Eq. (\ref{leptophobax}), namely the one where the dark sector contains the following $SU(2)_L\times U(1)_Y\times U(1)_{Y'}$ (fermionic) representations:
\bea
&&\chi_L\ (\ 1,\ \ \ \,0,\ \ \ \frac{3}{2}\ ) ,
\nonumber\\
&&\chi_R\ (\ 1,\ \ \ \,0,\ -\frac{3}{2}\ ) ,
\nonumber\\
&&\psi_L\ (\ 2,\ -\frac{1}{2},\ -\frac{3}{2}\ ) ,
\nonumber\\
&&\psi_R\ (\ 2,\ -\frac{1}{2},\ \ \ \frac{3}{2}\ ) ,
\nonumber\\
&&\eta_L\ (\ 1,\ \ -1,\ \ \ \frac{3}{2}\ ) ,
\nonumber\\
&&\eta_R\ (\ 1,\ \ -1,\ -\frac{3}{2}\ ) .
\label{bench}
\eea
In addition, the dark sector contains a complex scalar, $S$, with quantum numbers
\bea
\ \ \ \ \ \ S\ (\ 1,\ \ \ \,0,\ \ \ -3\ ) .
\eea
All the previous fields are color singlets. In the SM sector, only the quarks have non-vanishing $Y'$ charge: $Y'=1/3$.
 The model defined in Eq. (\ref{bench}) belongs to a class of leptophobic models formulated in Refs. \cite{Duerr:2013dza, Perez:2015rza}, from which we have borrowed the notation. The specific charge-assignment (\ref{bench}) was explicitly considered in \cite{Duerr:2015vna}.
%
%\cite{Duerr:2014wra}, from which we have borrowed the notation. However, the model was not analyzed by the authors.

With the previous spectrum, the most general fermionic Lagrangian involving fields of the dark sector reads
\bea
{\cal L}_{\rm fer} &\supset& {\cal L}_{\rm kin} - y_1 \bar \psi_L H \eta_R- y_2 \bar \psi_L \bar H \chi_R
- y_3 \bar \psi_R H \eta_L - y_4 \bar \psi_R \bar H \chi_L
\nonumber\\
&-&\lambda_\psi \bar \psi_L \psi_R S - \lambda_\eta \bar \eta_R \eta_L S - \lambda_\chi \bar \chi_R \chi_L S
- \lambda_L  \chi_L \chi_L S - \lambda_R  \chi_R \chi_R S^\dagger
\nonumber\\
&+& (\rm {h.c.}) .
\label{Lfer}
\eea
Similarly, the scalar Lagrangian involving the $S$ field is given by
\bea
{\cal L}_{\rm scal} &\supset& {\cal L}_{\rm kin} - m_S^2|S|^2 -\lambda_S^2|S|^4 - \lambda_{HS}^2|H|^2|S|^2 .
\label{Lscal}
\eea
Defining $S = \langle S \rangle + s$, the three parameters of Eq. (\ref{Lscal}) can be traded by $\langle S\rangle$, $m_s$ and the mixing between the Higgs boson and the scalar singlet $s$. This mixing is constrained by Higgs measurements. For the sake of simplicity, we will take $\lambda_{HS}=0$, so that there is no such mixing.

Notice that, even though the models in Eq. (\ref{bench}) with hypercharges $Y_\psi=\pm \frac{7}{2}$ and  $Y_\eta=\pm 5 $ have identical $Y'$ charges than the one we are considering, with this minimal particle content (3 fermions, the complex scalar $S$ and the gauge boson $Z^\prime$) they cannot be suitable DM models since the particular choice of hypercharges forbid operators coupling different dark fermions, like the ones in the first line of the Eq. (\ref{Lfer}). Thus an accidental flavour symmetry arises and the electrically charged  fermions, $\psi$, $\eta$, become stable.
%, cannot decay into the DM, becoming stable. 
This shortcoming might be avoided by enlarging the scalar sector with an extra Higgs with $Y_{H^\prime}=\pm \frac{3}{2}$. 
Consequently, the model defined in Eqs. (\ref{bench}-\ref{Lscal}) is somehow the minimal model with a
leptophobic $Z'$ mediator, axially coupled to the dark matter.
%Esto no me queda claro del todo si seria posible. Al final el extra Higgs tiene que decaer a SM porque si fuese estable seria tambien DM pero realmente no veo que tipo de acoplos puedes construir con este supuesto higgs y que no violen alguna simetria. Si no se puede quitar la frase del enlarge the sector.
  
Concerning the fermionic Lagrangian (\ref{Lfer}), it should be noticed that the ``Majorana couplings", $ \lambda_L,  \lambda_R$, if sizable, lead to the mixing and splitting of the two lightest degrees of freedom in the dark sector, so that the coupling of the lightest dark particle (i.e. the dark matter) to the $Z'$ would not be purely axial. This problem is avoided by noticing that taking $ \lambda_L=\lambda_R=0$, leads to a global $U(1)$ symmetry in the dark sector, under which all the dark fermions, $\{\chi, \psi, \eta\}$, transform with the same charge. This works exactly as a ``dark leptonic number". Consequently, we will assume  such global symmetry,  and thus $ \lambda_L=\lambda_R=0$.  (This assumption was not done in Ref. \cite{Duerr:2015vna}, so the model became non-axial.)

The extra fermionic fields in the dark sector, $\psi$ and $\eta$, can have an interesting phenomenology in colliders since they are charged under the SM gauge group. Furthermore, if they are light enough, they can play a relevant role in the dark matter phenomenology, in particular its thermal production in the early universe. E.g. if their masses are close enough to the DM one, their presence trigger efficient co-annihilation processes with the DM particle. However, since we are interested in exploring characteristics  of the simplest scenario, we will make the assumption that the $\psi$ and $\eta$ masses are large enough to integrate these fields out. 
 In that regime we recover a scenario which is similar to SDMM, but with some differences, e.g.  the correlation between the coupling of the $Z^\prime$ to the SM and dark fields (which are taken as free parameters in SDMM). In this way, we get a truly realistic a SDMM (as it emerges from an anomaly-free model), whose performance is worth to examine. As we are about to see, even in that case, the extra fields leave a footprint in the low-energy theory in the form of an effective operator.  The present analysis can be thus considered as the study of a portion of the parameter space of the theory, but of course the remaining regions are also interesting and would require a specific study.

On the other hand, the `dark scalar', $s$, may play a relevant role in DM annihilation at the early universe, due for instance to the $s-$channel process $\chi\chi\rightarrow s\rightarrow Z' Z'$. Depending on the values of $m_\chi, m_s$, this diagram can be competitive with the diagram $\chi \chi \rightarrow Z' Z'$, where $\chi$ propagates in $t-$channel. (Both diagrams are shown in see Fig. \ref{fig:diagrams1} below.) Actually, for $m_s\sim 2\ m_\chi$ the $s-$mediated annihilation becomes resonant and dominant (`$s-$funnel'). The effect of the $s-$field in the DM phenomenology has been discussed in Ref. \cite{Duerr:2016tmh}.
Along the paper we will consider two possibilities, namely a heavy scalar, $m_s^2\gg m_\chi^2$, and a not-too-heavy one, in order to show its impact on the DM physics and phenomenological prospects.

Hence, after integration of the extra dark fermions, we end up with an effective theory where the dark sector contains just the DM field, $\chi$, besides the scalar $s$ and the $Z'$ mediator. In addition there is an effective Dim$-$5 operator, $\sim |H|^2 \bar\chi_L \chi_R$, which arises upon the integration of $\psi$ field. Thus, the relevant DM Lagrangian of the effective theory reads

\bea
{\cal L}_{\rm eff} ^{\rm DM}= 
{\cal L}_{\rm kin}  -  \lambda_\chi  \bar \chi_R \chi_L S + \frac{1}{\Lambda}  \bar \chi_R \chi_L |H|^2 +\cdots
+ (\rm {h.c.}) ,
\label{Leff}
\eea
where it is understood that ${\cal L}_{\rm kin}$ contains the gauge interactions with the $Z'$ and 
\bea
\frac{1}{\Lambda} = \frac{y_2 y_4}{m_\psi} .
\label{Lambda}
\eea
Note that this operator is exactly the one of a fermionic singlet Higgs$-$portal. Therefore, a $Z'-$framework naturally leads to a Higgs$-$portal, thus representing an interesting UV completion of it. Nevertheless this ``Higgs$-$portal" operator
%
%{\blue comentar incidentalmente que este operador es un fermionic Higgs-portal. Por tanto este esquema de una Z' conduce naturalmente a un Higgs portal, siendo una interesante UV compltion del mismo. Sin embargo...}
%
is not going to play any relevant role in the DM phenomenology. The reason is that if the effective coupling $1/\Lambda$ is large enough to contribute to the DM annihilation in the early universe, 
%through the $\chi \chi \rightarrow h h$ process, 
then the strong constraints from direct (and indirect) detection rule out the scenario in most of the parameter space (except very close to the Higgs$-$funnel, $m_\chi \simeq m_h/2$). This will be discussed below. Consequently, we will assume in (most of) what follows that $1/\Lambda$ is small enough to be neglected.

In this regime, the model is thus described by three parameters: the $U(1)_{Y'}$ gauge coupling\footnote{The notation $g_B$ stems from the equivalence of $U(1)_{Y'}$ and $U(1)_B$ for the SM fields.}, $g_B$; the $Z'-$mass, $m_{Z'}$ (or, equivalently, $\langle S\rangle$); and the dark matter mass, $m_\chi\simeq \lambda_\chi \langle S\rangle$. 
In the case of a not-too-heavy $s-$field, there is one extra relevant parameter, $m_s$ (the coupling of $s$ to $\chi\chi$, $\lambda_\chi$, is determined by the value of $m_\chi$). 
This is to be compared with ordinary SDMM, 
where there are four parameters, since the gauge coupling of the $Z'$ to quarks ($g_q$) and to DM ($g_{\rm DM}$) are taken as independent parameters. As explained above, in our scenario, the cancellation of anomalies fixes the ratio between them: $g_{\rm DM}/g_q = 4.5$.

Still, we will see that the model is perfectly viable and quite predictive.

%%%%%%%%%%%%%%%%%%%%%%%%%%%%%%%%%%%%%%
\section{Phenomenology of the Model}
\label{sec:pheno}
%%%%%%%%%%%%%%%%%%%%%%%%%%%%%%%%%%%%

%%%%%%%%%%%%%%%%%%%%%%%%%%%%%
\subsection{Kinetic mixing}
%%%%%%%%%%%%%%%%%%%%%%%%%%%%%

As it is well known, the presence of more than one $U(1)$ factor in the gauge group leads to the possibility of kinetic terms which mix the corresponding gauge fields. In our case, such kinetic-mixing term takes the form
\bea
{\cal L}_{\rm kin} \supset -\frac{1}{2}\ \epsilon\ F_{\mu \nu}^Y\ F^{Y' \mu \nu} .
\label{Lkinmix}
\eea
where $F^{Y (Y')}$ is the field-strength tensor of the $U(1)_{Y(Y')}$ gauge factor.

It is reasonable to assume that $\epsilon=0$ at some unknown high-energy scale, $\Lambda'$, above which the theory enters a different ultraviolet regime. Still, since quarks couple to both $U(1)$ gauge bosons, quark loops generate a non-vanishing value of $\epsilon$ at lower energies, $\mu = m_{Z'}$ \cite{Carone:1995pu}
\bea
 \epsilon=\frac{e g_q}{2 \pi^2\cos \theta_W} \log \frac{\Lambda'}{\mu} \simeq 0.02\  g_q \ \log\frac{\Lambda'}{\mu} ,
\label{epsilon}
\eea
where $g_q=g_{Y'}/3$. Note that this result is completely general for any leptophobic model since, as commented in Sec. \ref{sec:model}, leptophobia implies that $U(1)_{Y'}$ is equivalent to baryon number for the SM fields. 
In addition to quarks, there are loops involving the $\eta$, $\psi$ fields, which are also charged under both $U(1)$s. However, the fact that their coupling to $U(1)_Y$ ($U(1)_{Y'}$) are vectorial (axial) makes their contributions to $\epsilon$ to cancel. In consequence, Eq.(\ref{epsilon}) holds.
The previous mixing leads to relevant phenomenological constraints, e.g. from electroweak (EW) observables and di-lepton production at the LHC, which will be discussed in Sec. \ref{sec:results}.

In order to prepare the model for the phenomenological analysis, one has to properly normalize and diagonalize the gauge kinetic terms.
 We have followed here the analysis of Refs. \cite{Kahlhoefer:2015bea, Coriano:2015sea}. To summarize, after appropriate redefinition of the $U(1)_{Y'}$ gauge boson,  the kinetic terms get diagonal and normalized, while the covariant derivative takes the form
\bea
{\cal D}_\mu=\partial_\mu + i g_s T^aG^a_\mu + i g t^aW^a_\mu +i g' Y B_\mu + i(\tilde gY +g_{B}Y')B'_\mu .
\label{CovDer}
\eea
where $G_\mu, W_\mu, B_\mu$ are the ordinary gauge bosons of $SU(3)_c\times SU(2)_L\times U(1)_Y$; $B_\mu'$ is the gauge boson of $U(1)_{Y'}$ (with a small admixture of $B_\mu$) and
\bea
\tilde g = \frac{\epsilon}{\sqrt{1-\epsilon^2}}\ g' \simeq \epsilon g' .
\label{gtilde}
\eea
The final physical fields, $A_\mu, Z_\mu, Z'_\mu$, are obtained upon diagonalization of the gauge-boson mass matrix:
\bea
\left(
{\begin{array}{c}
   B_\mu  \\
   W^3_\mu  \\
   B'_\mu
  \end{array} }
\right) =
\left(
{\begin{array}{ccc}
   \cos\theta_w & -\sin\theta_w \cos\theta'& \sin\theta_w \sin\theta' \\
    \sin\theta_w& \cos\theta_w \cos\theta' & -\cos\theta_w \sin\theta' \\
   0 & \sin\theta' & \cos\theta' 
  \end{array} }
  \right)
  \left(
{\begin{array}{c}
   A_\mu  \\
   Z_\mu  \\
   Z'_\mu
  \end{array} }
\right) ,
\label{gaugematrix}
\eea
where $\theta_w$ is the weak angle and $\theta'$ is the mixing between the $Z$ and $Z'$ fields, given by\footnote{Eq. (\ref{thetap}) is accurate enough for small $\epsilon$; the complete expression can be found e.g. in ref.\cite{Coriano:2015sea}, Eq. (44).}
\bea
\theta' \simeq \epsilon\sin\theta_w\frac{m_Z^2}{m_{Z'}^2-m_Z^2} . 
\label{thetap}
\eea
All these relations will be applied below. %in subsection XX below.

%%%%%%%%%%%%%%%%%%%%%%%%%%%%%%%%%%%%%%%%%%%%%%%%%%%%%%%%%%%%%%%%%%%%%
\subsection{Dark Matter Constraints}
%%%%%%%%%%%%%%%%%%%%%%%%%%%%%%%%%%%%%%%%%%%%%%%%%%%%%%%%%%%%%%%%%%%%

\begin{figure}[h!]
\centering 
\includegraphics[width=0.4\linewidth]{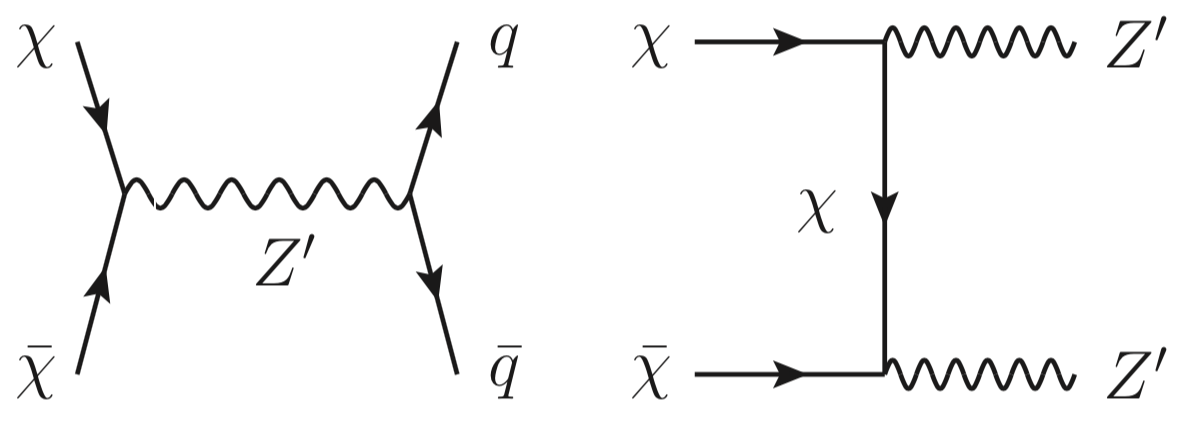}
\includegraphics[width=0.6\linewidth]{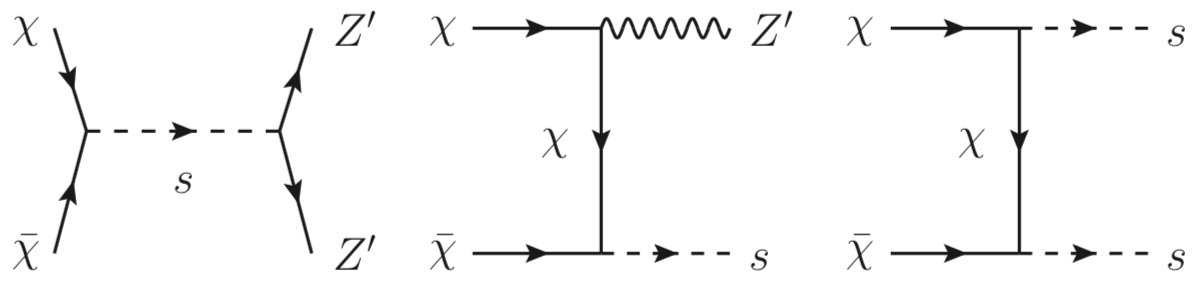}
\caption{Feynman diagrams, 
%appearing at the Lagrangian level, 
relevant for DM annihilation in the model.}
\label{fig:diagrams1}
\end{figure}

\begin{figure}[h!]
\centering 
\includegraphics[width=0.4\linewidth]{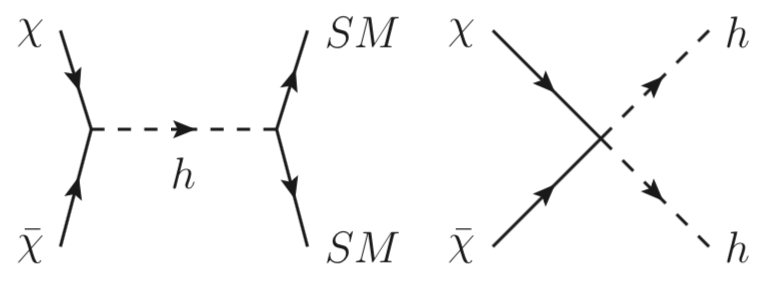}
\caption{Feynman diagrams arising from the effective operator (\ref{Leff}), that contribute to  DM annihilation in the model.}
\label{fig:diagrams2}
\end{figure}

From the Lagrangian of the model (\ref{Leff}), the thermal production of dark matter in the early universe is controlled by the DM annihilation processes of Figs. \ref{fig:diagrams1}, \ref{fig:diagrams2}.

Keeping for the moment the assumption that the effective coupling, $1/\Lambda$, in Eq.(\ref{Leff}) is small (which is perfectly reasonable), the main annihilation channels of DM come from the first two diagrams
of Fig.~\ref{fig:diagrams1} (and the other three as well if $s$ is light enough).
Thus the annihilation rate depends on the main three parameters of the model, $\{g_B, m_{Z'}, m_\chi\}$ (plus $m_s$ if the $s-$field is relevant). Recall that
%, if the particle content of the dark sector is minimal, the requirement of leptophobia and axial DM-coupling  fixes 
the relative couplings of $Z'$ to quarks and DM are determined by $g_B$, namely $g_q = \frac{1}{3}g_B$, $g_{\rm DM}= \frac{3}{2}g_B$. Consequently, for each value of $\{m_{Z'}, m_\chi, m_s\}$, there is always a (unique) value of $g_B$ (maybe in the non-perturbative regime) which leads to the correct relic DM density, $\Omega_{\rm DM} h^2 =0.1188$ \cite{Ade:2015xua}.

\begin{figure}[h!]
\centering 
\includegraphics[width=0.49\linewidth]{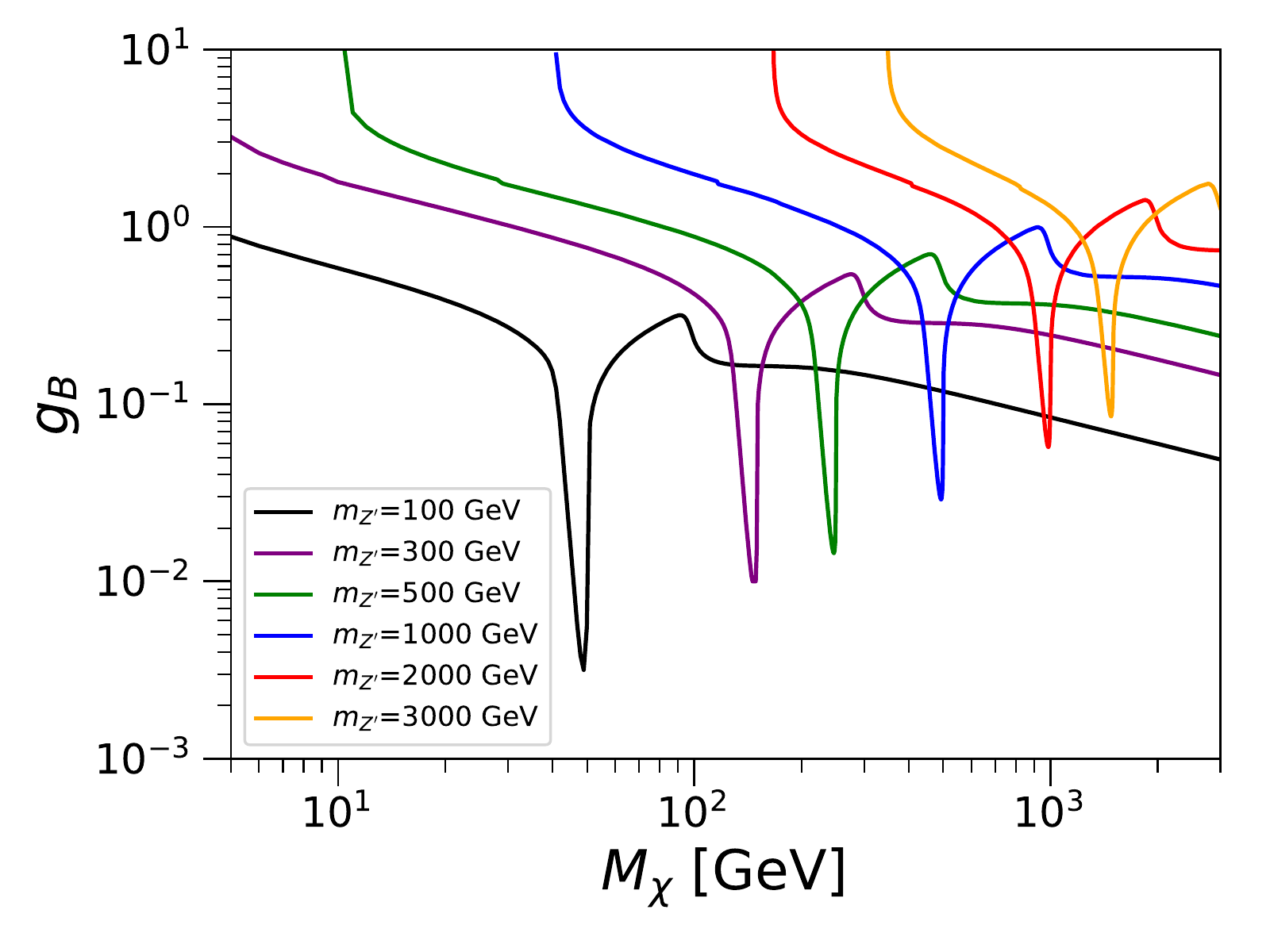}
\includegraphics[width=0.49\linewidth]{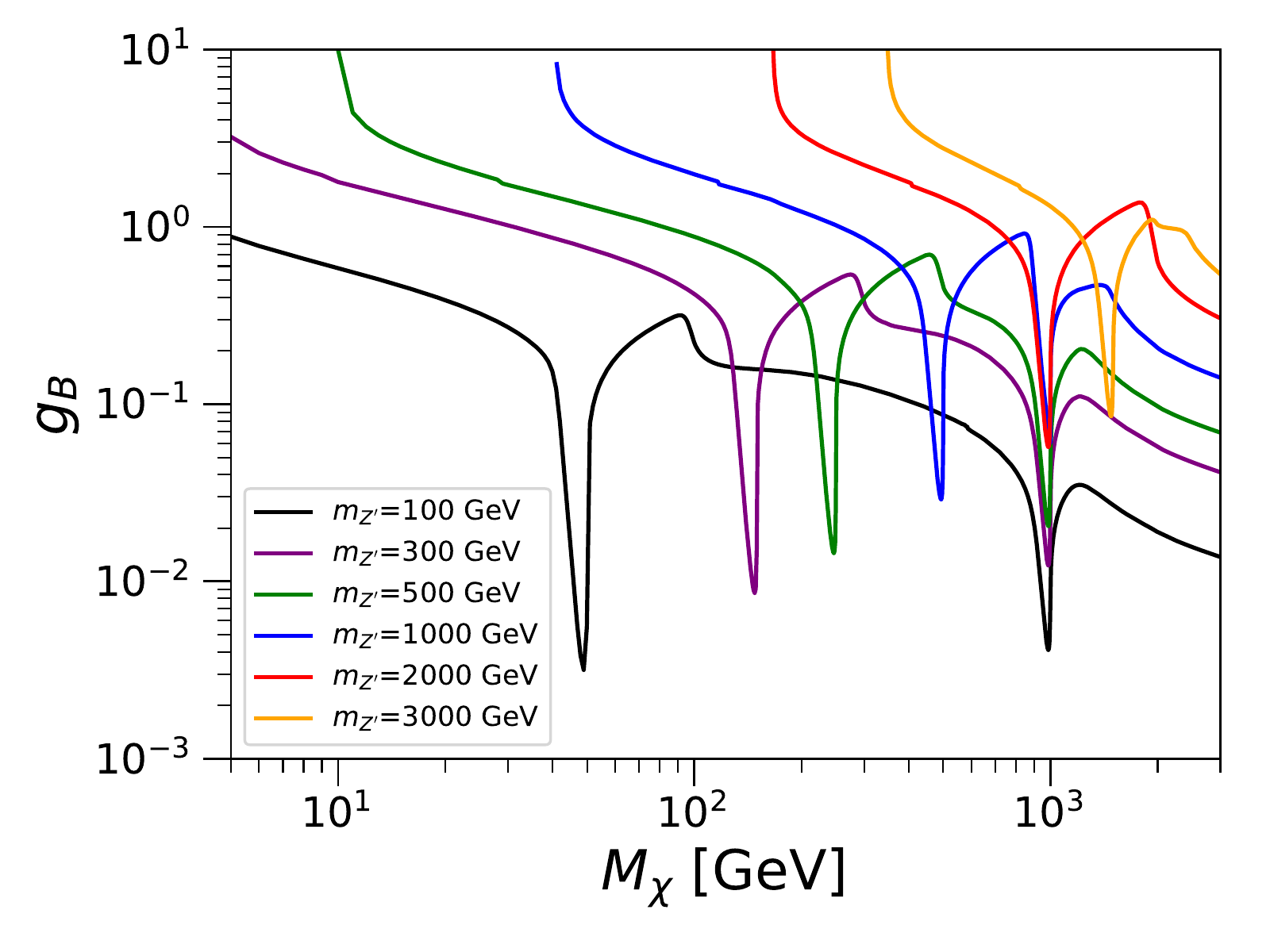}
\caption{
Values of the $g_B$ coupling that reproduce the observed DM relic density as a function of the DM mass for several choices of $m_{Z'}$. The left (right) panel shows the  $m_s = 15$ TeV  ($m_s = 2$ TeV) case.
}
\label{fig:mXgB}
\end{figure}

This is illustrated in Fig. \ref{fig:mXgB} in the $m_\chi-g_B$ plane for several choices of $m_{Z'}$ and two choices of the scalar mass, $m_s = 15$ TeV (i.e. irrelevant) and $m_s = 2$ TeV. Interestingly, the value of $g_B$ remains in the perturbative regime in most of the parameter space. For each curve, the two resonances, $2m_\chi\sim m_{Z'}, m_s$, and the threshold of two $Z'$s are visible. Note that the values of $g_B$ are almost the same in both panels, unless $m_s \lsim 2m_\chi$, i.e. when the effects of the scalar in the DM annihilation are non-negligible.

Concerning bounds from direct and indirect detection, as mentioned in previous sections, the fact that the $Z'$ couples to DM (SM quarks) in an axial (vectorial) way, implies that the effective DD interaction is spin-dependent and velocity-suppressed \cite{Arcadi:2014lta}. Analogously, indirect detection (ID) is velocity-suppressed as well \cite{Arcadi:2014lta}.
Consequently, there are virtually no bounds from DD or ID on the model (for $1/\Lambda$ small). Actually, the most important constraints on the model (and the opportunity to probe it experimentally) come from collider measurements, which we examine in the next subsections.

\begin{figure}[h!]
\centering 
\includegraphics[width=0.6\linewidth]{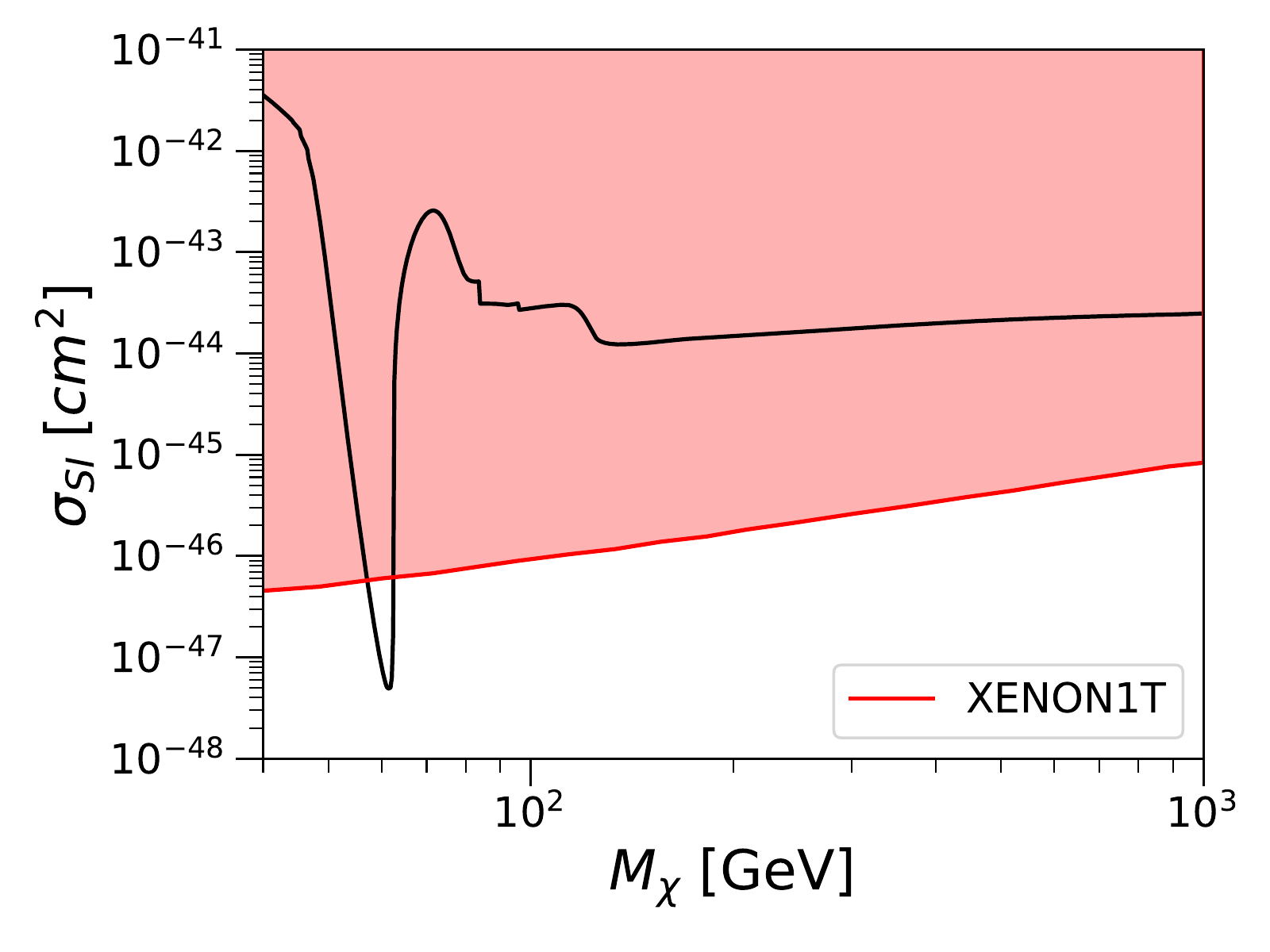}
\caption{DM-nucleon spin-independent cross-section as a function of the DM mass when DM annihilation occurs thanks to the effective operator of Eq. (\ref{Leff}). The black line corresponds to the observed relic density. The red-shared area is excluded by current XENON1T constraints. 
}
\label{fig:DMeffop}
\end{figure}

Let us finish this subsection by discussing the role of the effective ``Higgs$-$portal" operator of Eq. (\ref{Leff}) in the DM phenomenology. This interaction leads to the DM annihilation processes of Fig. \ref{fig:diagrams2}.
In Fig. \ref{fig:DMeffop} we have plotted (black line) the corresponding spin-independent DM-nucleon cross section as a function of $m_\chi$ when the value of the effective coupling, $1/\Lambda$, is adjusted to reproduce the relic density; showing as well the region excluded by the current XENON1T limits \cite{Aprile:2018dbl}.
Only a narrow range of $m_\chi$ around the Higgs-funnel region is still surviving. Hence the effective Higgs$-$portal operator must be suppressed enough to avoid these strong bounds (fortunately this is perfectly sensible from (\ref{Lambda})), and it is reasonable to assume that all the DM annihilation occurs through the diagrams of Fig. \ref{fig:diagrams1}.
%
%%%%%%%%%%%%%%%%%%%%%%%%%%%%%%%%%%%%%%%%%%%%%%%%%%%%%%%%%%%%%%%%%%%%%%%
\subsection{Bounds from EW observables and LHC}
%%%%%%%%%%%%%%%%%%%%%%%%%%%%%%%%%%%%%%%%%%%%%%%%%%%%%%%%%%%%%%%%%%%%%%%

As mentioned above, the presence of a kinetic mixing, $\epsilon$,
%(\ref{Lkinmix}) 
between the two $U(1)$ gauge groups is unavoidable due to radiative corrections involving quarks. In the following we will assume that $\epsilon$ is initially vanishing at some unknown UV scale, $\Lambda'$, so that its effective value at the $m_{Z'}$ scale is given by Eq. (\ref{epsilon}). We will derive results for two representative choices of the UV scale: $\log(\Lambda'/ m_{Z'})= 1,\ 4.6$ (the latter corresponds to $\Lambda' =100\ m_{Z'}$).

A non-vanishing $\epsilon$ induces important physical effects which constrain the model. The most relevant ones are electroweak precision observables, EWPO, particularly, $S$ and $T$, and the production of di-leptons at the LHC.

Concerning the first ones, we use the well-known expressions for the oblique parameters $S$ and $T$ \cite{Kahlhoefer:2015bea} 
\bea
\alpha_{\rm em}\ S&=&4 c_w^2 s_w \theta'\left( \epsilon-s_w\theta'\right) ,
\nonumber\\
\alpha_{\rm em}\ T&=&\theta'^2\left(\frac{m_{Z'}^2}{m_Z^2}-2\right)+
2 s_w \theta' \epsilon ,
\label{alphaST}
\eea
and take $S = 0.03 \pm 0.10$, $T = 0.05 \pm 0.12$ as values derived from the global fit to the electroweak precision data performed in Ref. \cite{Baak:2013ppa}. 

We recall that the mixing angle $\theta'$ involved in Eq. (\ref{alphaST}) is given in terms of $\epsilon$ and $m_{Z'}$ by Eq. (\ref{thetap}). Obviously, for a given $\epsilon$, the larger $m_{Z'}$ the smaller $\theta'$. Consequently, EW observables can be relevant at small $m_{Z'}$. 

Regarding di-leptons, the kinetic mixing triggers couplings of the $Z'$ to leptons, as it is clear from Eqs. (\ref{CovDer}), (\ref{gaugematrix}) (the precise expressions for the couplings to $\ell_L$, $\ell_R$ leptons can be found in Refs. \cite{Coriano:2015sea, Kahlhoefer:2015bea}). Hence, production of $Z'$s at the LHC leads to the possibility of di-leptons at the final state.
LHC has provided strong constraints on the di-lepton search using $36.1$ fb$^{-1}$ data at $\sqrt{13}$ TeV. Ref. \cite{ATLAS-CONF-2017-027} gives bounds on the coupling of $Z'$ to leptons as function of $m_{Z'}$ for several representative examples of the associated $U(1)_{Y'}$. More precisely, that reference provides an analysis on the bounds on a $Z'$ corresponding to $B-L$, which is identical to ours for quarks, and thus for $Z'$ production. Then the ratio of the braching fraction of $Z'$ into leptons in the $B-L$ model over the one in ours, can be straightforwardly derived from the respective couplings of both $Z'$s to leptons. In addition, it has to be taken into account that, depending on the value of $m_{Z'}$, the gauge boson can decay into top-antitop and/or $\chi\chi$ (with appropriate kinematical factors), which modifies further the braching fraction into leptons. We have taken into account all these details in order to extract the bounds from di-leptons, which will be shown in the next subsection. 

Bounds from di-leptons are stronger for smaller $m_{Z'}$. Hence, as for EWPO, the constraints on our model due to kinetic mixing are specially relevant in the range of light $Z'$. Needless to say, the larger the UV scale, $\Lambda'$, the larger the radiatively induced $\epsilon$ and thus the stronger both types of bounds. 

Constraints from di-jet searches turn out to be the dominant ones in most of the parameter space. We have translated the last ATLAS results on di-jets \cite{ATLAS-CONF-2016-070, Aaboud:2017yvp, Aaboud:2018fzt, Aaboud:2018tqo, Aaboud:2018zba} into bounds on the scenario at hand. 
As for the above di-lepton bounds, this entails to take into account that, depending on the value of $m_{Z'}$, the gauge boson can decay into top-antitop and/or $\chi\chi$ (with appropriate kinematical factors), thus modifying the branching fraction into di-jets. 
In the $m_{Z'} \sim 140-500$ GeV mass window, where UA2 \cite{Alitti:1993pn} and CDF \cite{Aaltonen:2008dn} experiments have better sensitivity than LHC experiments, the limits are however weaker than mono-jet bounds, which are discussed next.

Finally, mono-jet production at the LHC from ISR in the $q\bar q\rightarrow Z'\rightarrow \chi\chi$ process leads to important constraints on the model, which are specially relevant in the region of light $Z'$. This type of signatures are characterized by a high-pT object recoiling against $\cancel{\it{E}}_{T}$ which can be triggered at the ATLAS and CMS detectors. Our application of the mono-jet constraints is based on its implementation in MicrOMEGAS \cite{Barducci:2016pcb}, with 20.3 fb$^{-1}$ data collected at $\sqrt{8}$ TeV \cite{Aad:2015zva} \footnote{We have checked that the coverage of current 13 TeV data is similar.}.

%%%%%%%%%%%%%%%%%%%%%
\section{Results}
\label{sec:results}
%%%%%%%%%%%%%%%%%%%%%

We have scanned the DM mass and $Z'$ mass plane randomly for two different values of the scalar s-field mass ($m_s = 2, 15$ TeV) requiring each point to fulfill the central value of the Planck measured DM relic density $\Omega h^2 = 0.1188$ \cite{Ade:2015xua}.
%{\red SC: I assume you included some uncertainty margin here ? RR: No really we require to reproduce exactly the mean value. Since the uncertainties are small it does not matter in terms of allowed values of the coupling.}
This procedure fixes the coupling $g_B$. Besides, we impose a $2 \sigma$ cut on the $S$ and $T$ oblique parameters and apply 95\% C.L. exclusion limits from LHC searches of di-leptons, di-jets and mono-jets as it has been discussed in Sec. \ref{sec:pheno}.

For the calculation of the relic density the program MicrOMEGAS \cite{Barducci:2016pcb} has been used. MicrOMEGAS is based on the CalcHEP \cite{Belyaev:2012qa} package which is used to calculate the tree level cross sections relevant for DM annihilations and thus the DM relic density. The implementation of the model in CalcHEP format has been done using the FeynRules package \cite{Alloul:2013bka}.

\begin{figure}[h!]
\centering 
\includegraphics[width=0.49\linewidth]{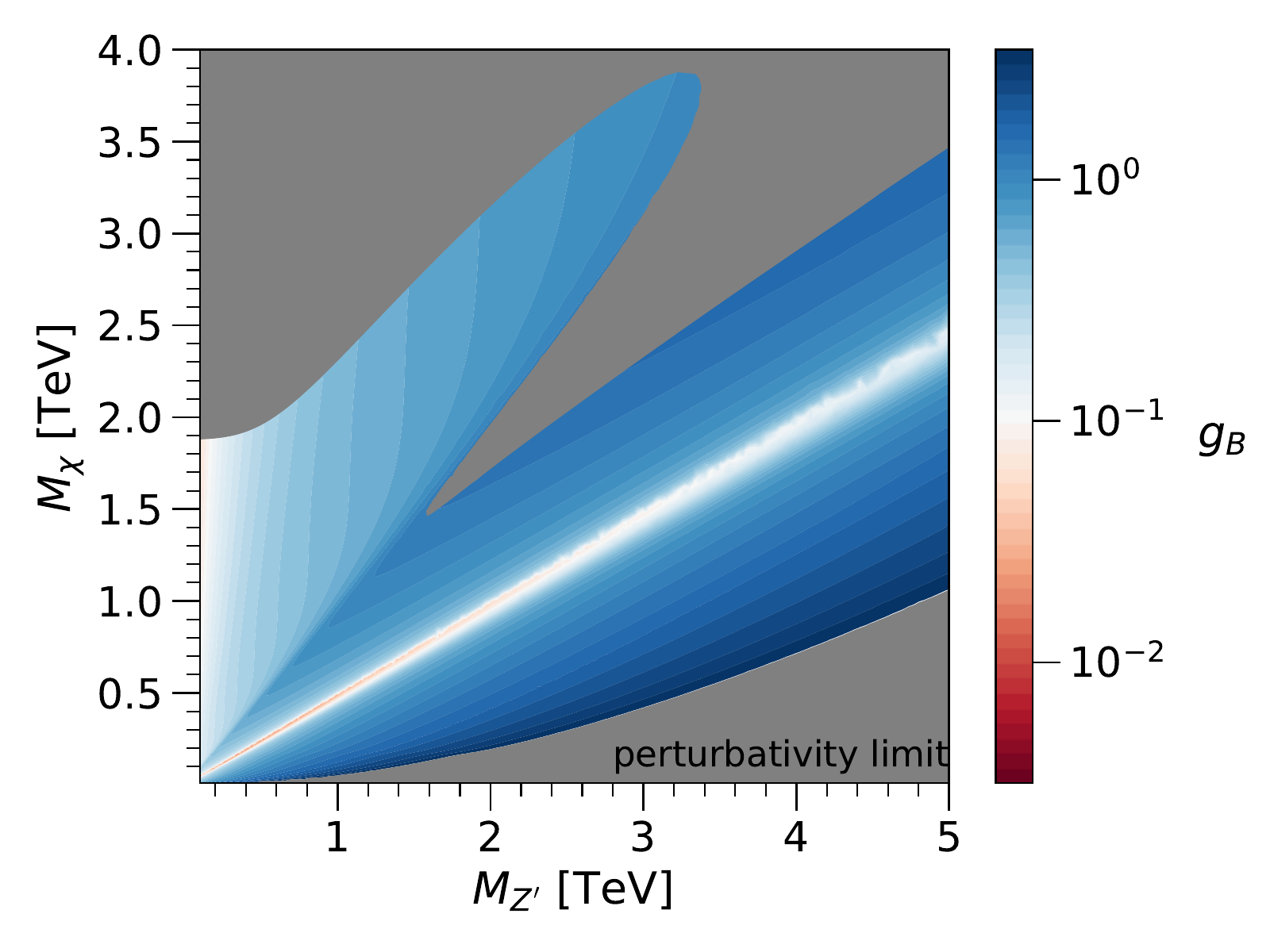}
\includegraphics[width=0.49\linewidth]{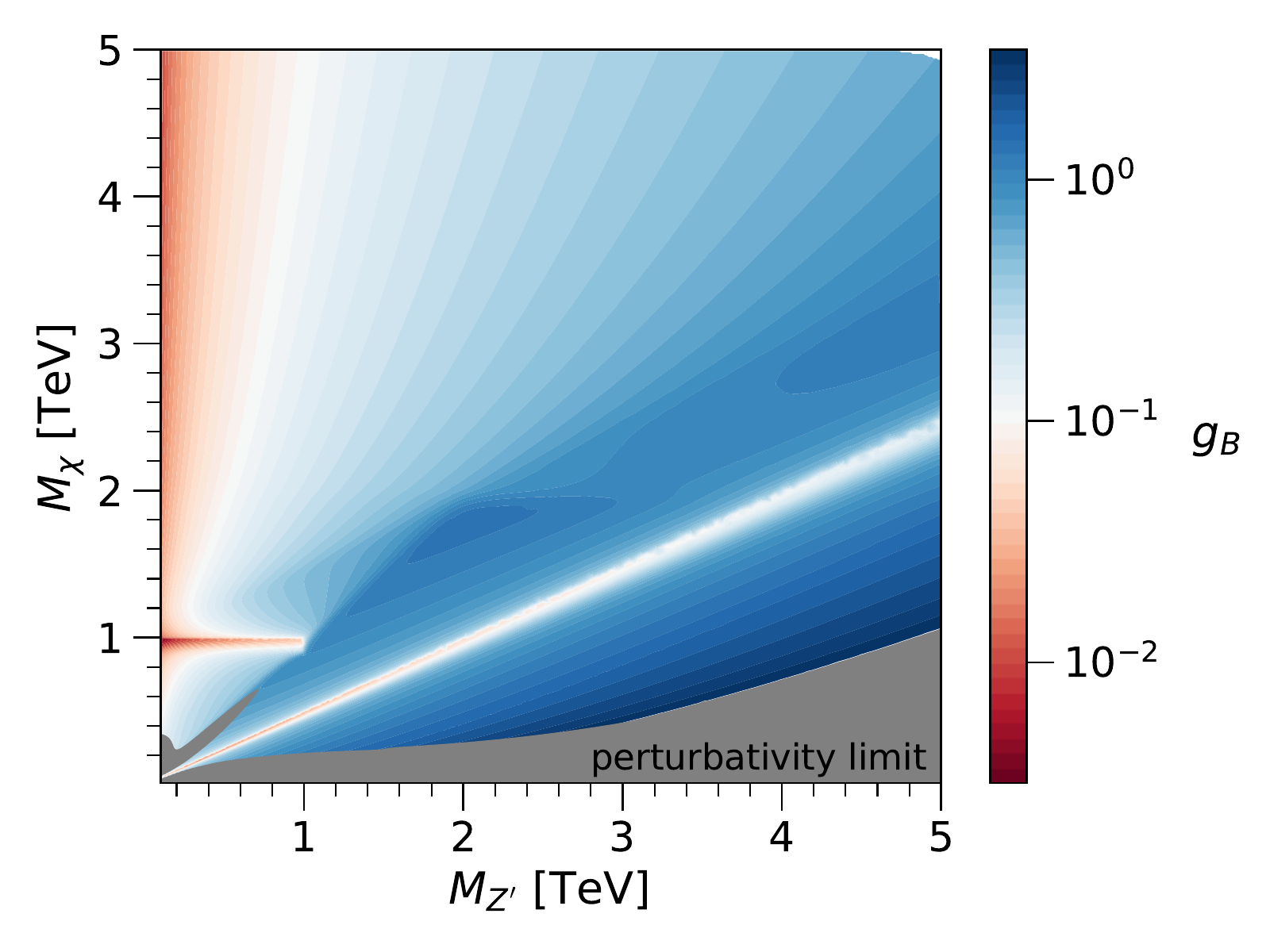}
\caption{
The logarithmic-scale colorbar gives the values of the $g_B$ coupling that fit the observed DM relic density in the $m_{Z'}- m_\chi$ plane. The left (right) panel shows the $m_s > 2m_\chi$  ($m_s = 2$ TeV) case. The grey-shaded region is excluded by the perturbativity condition in the various couplings.
}
\label{fig:gB}
\end{figure}

As explained in previous subsections, our model, which is representative of a leptophobic $Z'$ axially coupled to DM with minimal dark sector, has only three relevant parameters: $\{g_B, m_{Z'}, m_\chi\}$, plus $m_s$ if the scalar is not too heavy. We have considered here the simplest possibility where effective interactions due to the extra dark fermions, $\psi$ and $\eta$, are negligible since their masses are substantially bigger than $m_{Z'}, m_\chi$. The study of phenomenological implications of these extra dark fermions is left for a future work.
%{\red SC: The dark fermions are phenomenologically interesting and new, can one add something about the possible LHC search channels ? RR: We have commented about this just above and in the conclusions}
It was shown in Subsec. 4.2 that for any choice of $m_{Z'}, m_\chi, m_s$, there is a unique value of $g_B$ leading to the correct thermal relic density, $\Omega_{\rm DM} h^2$. Fig. \ref{fig:gB} shows such value of $g_B$ in the $m_{Z'}- m_\chi$ plane for two regimes of $m_s$.  In most of the interesting parameter space $g_B$ is well inside the perturbative regime, which we have taken as $g_B < 4\sqrt{\pi}$ (see \cite{Hinchliffe} for a detailed discussion). However, the most important restrictions from the perturbativity requirement come from the fermionic Yukawa couplings, $\lambda_{\chi,\psi,\eta}$, and, the scalar one, $\lambda_S$. The latter is the most constraining one in the regime where $m_s>2m_\chi$ (left plot of Figure \ref{fig:gB}), i.e. when the scalar plays a negligible role for the DM annihilation in the early universe. In contrast, when the scalar plays a role ($m_s\lsim 2m_\chi$), the required value of $g_B$ becomes smaller. This is illustrated in the right plot of Figure \ref{fig:gB} for $m_s=2$ TeV.
In consequence, for a given value of $m_{Z'}$, the VEV $\langle S\rangle$ becomes larger and all the (fermionic and scalar) couplings smaller. Then, the perurbative limits exclude a much smaller region in the parameter space, as shown in the figure.
%{\red Recall also that the coupling of $Z'$ to quarks is smaller, namely $g_q= g_B/3$.} 
The resonance region, $2m_\chi\sim m_{Z'}, m_s$ is also visible in the figure.

The trend in both cases is that the larger (smaller) $m_\chi$ ($m_{Z'}$) the smaller $g_B$. As we shall see shortly, this will be, in general terms, the region safe with respect to the various constraints and, consequently, it becomes larger in the regime where the scalar field plays a significant role.

%relevant $\chi\chi\rightarrow Z'Z'$ {\magenta maybe we should include here the channels $\chi\chi\rightarrow Z's/ s s$ that are also very relevant (even dominant) in some cases when they are kinematically allowed. Here alse we can refer to the 3 processes in the low part of fig \ref{fig:diagrams1}, to illustrate this} is the dominant DM annihilation process and the $s-$mediated diagram is relatively more significant.

%
Next we show the phenomenological bounds on the model in the same $m_{Z'}- m_\chi$ plane, assuming at any point the value of $g_B$ leading to the correct $\Omega_{\rm DM}$, as given in Fig. \ref{fig:gB}. 

\begin{figure}[h!]
\centering 
\includegraphics[width=0.49\linewidth]{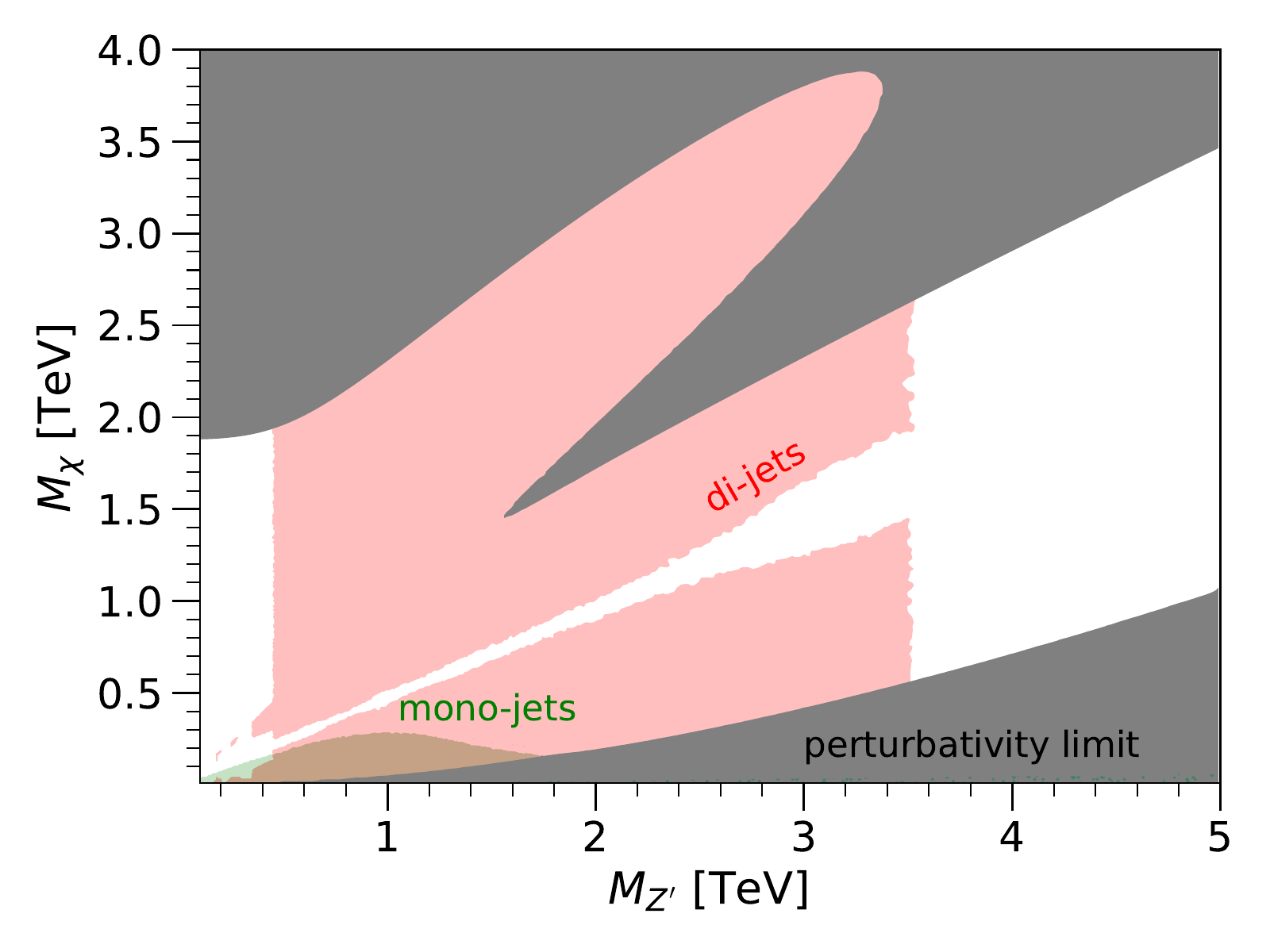}
\includegraphics[width=0.49\linewidth]{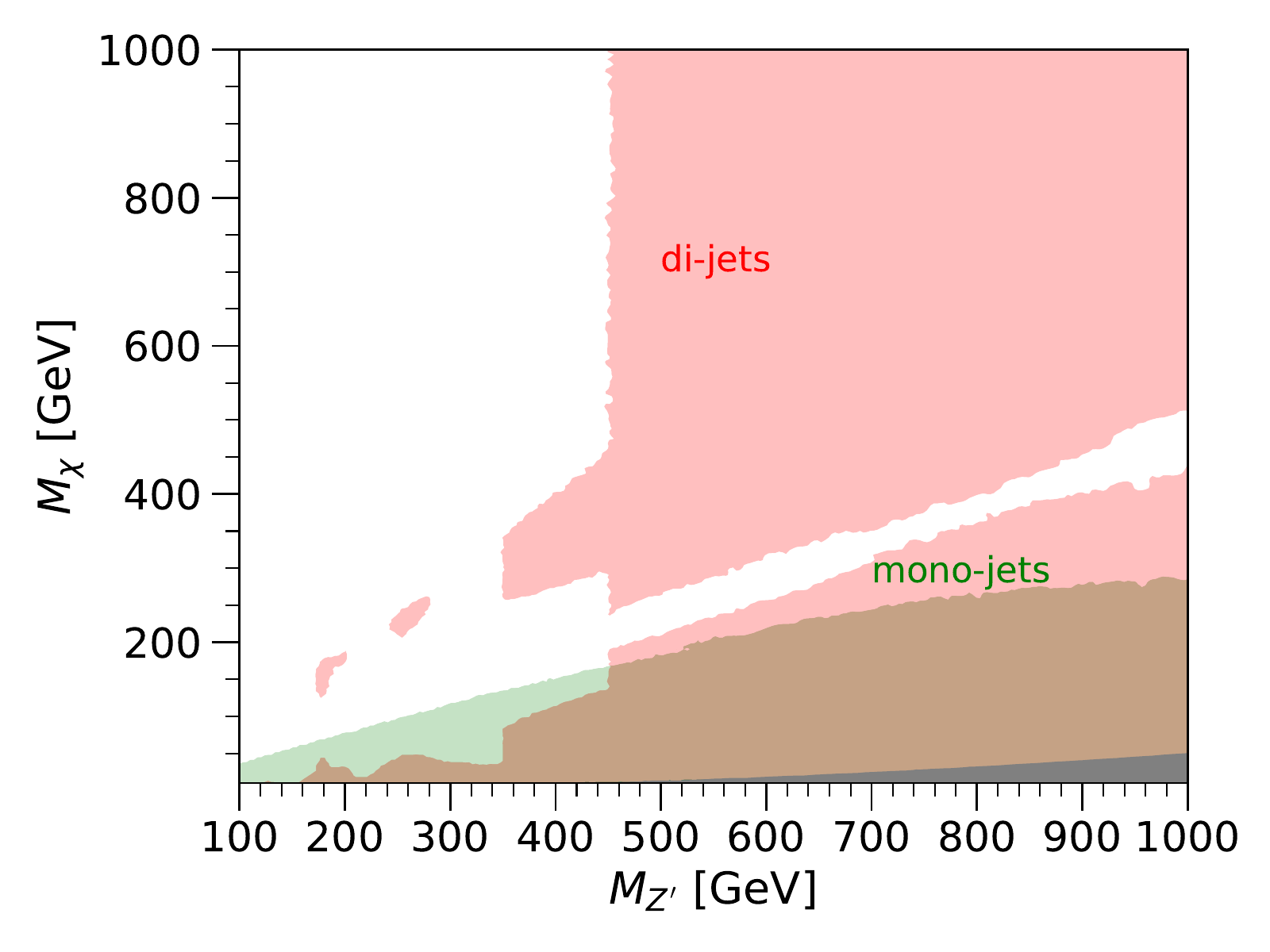}\\
\includegraphics[width=0.49\linewidth]{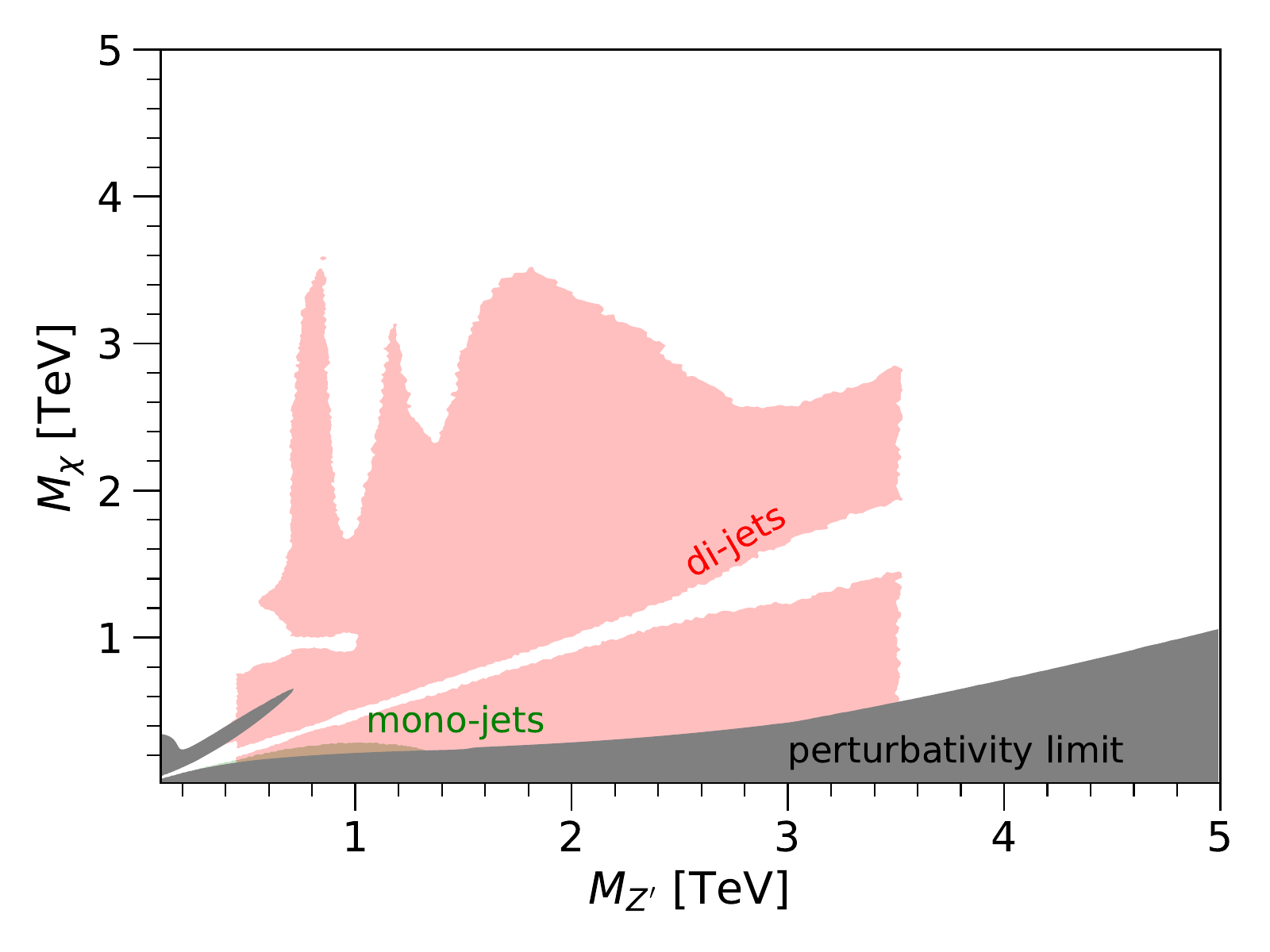}
\includegraphics[width=0.49\linewidth]{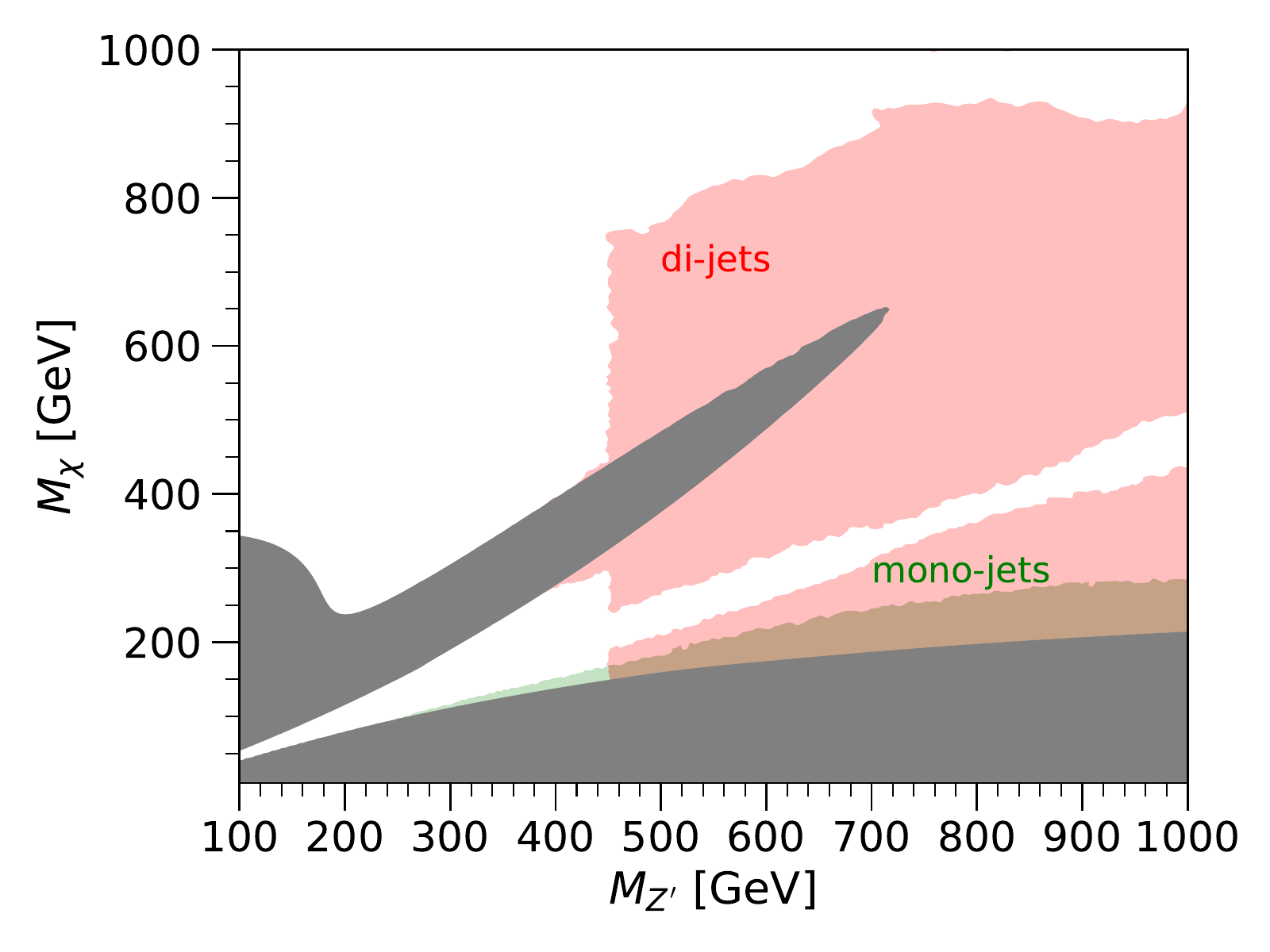}
\caption{
Areas in the $m_{Z'}- m_\chi$ plane forbidden by constraints from di-jets (pink) and mono-jets (green); for log $(\Lambda'/ m_{Z'})=1$ (see Eq.~\ref{epsilon}).
The value of the $g_B$ coupling is adjusted at every point to reproduce the observed DM relic density.
In the grey region the coupling becomes non-perturbative. Upper (lower) panels show the case where $m_s > 2 m_\chi$ ($m_s = 2$ TeV). Left panels show the full range of $m_{Z'}$ considered while in the right ones we zoom in the region of $Z'$ masses up 1 TeV.}
\label{fig:Lambdalow}
\end{figure}

Fig. \ref{fig:Lambdalow} shows the constraints on the model discussed in the previous subsection for $\log (\Lambda'/ m_{Z'})=1$. As expected di-jet production (pink region) gives the dominant constraint in most of the parameter space. It essentially excludes the whole $500\ {\rm GeV}\simlt m_Z' \simlt 3000\ {\rm GeV}$ region, except around the $Z'$ and $s$ resonances, $2m_\chi\sim m_{Z'}, m_s$.
 Notice that the constraints from a correct relic density are also incorporated, as every point in the $m_{Z'}- m_\chi$ plane has the correct relic density,  according to Fig. \ref{fig:gB}.

%{\magenta En los graficos que tenemos esa garganta se corta en $m_s<2 m_{Z'}$, aunque esto no lo entiendo muy bien. Es cierto que a partr de esa masa de la $Z'$ no es posible producir dos $Z'$ on-shell a traves del proceso $\chi\chi\rightarrow s\rightarrow Z'Z'$. pero una de ellas puede estar off-shell y decaer en $q\bar q$ (es la unica forma que tiene de decaer). Supongo que lo que sucede es que ahora el escalar $s$ prefiere ir otra vez a $\chi\chi$, de forma que al final el proceso no aniquila materia oscura. Este efecto tambien depende del hecho de que hemos tomado un acoplo muy peque\~no (o cero) $\lambda_{HS}$, si no, $s$ puede decaer en $hh$} 
%However, we do not see such throat because we have assumed a larger mass for $s$. {\magenta No se si este punto del $s-$funnel es suficientemente interesante para ilustrarlo en un plot}.
%

\begin{figure}[h!]
\centering 
\includegraphics[width=0.49\linewidth]{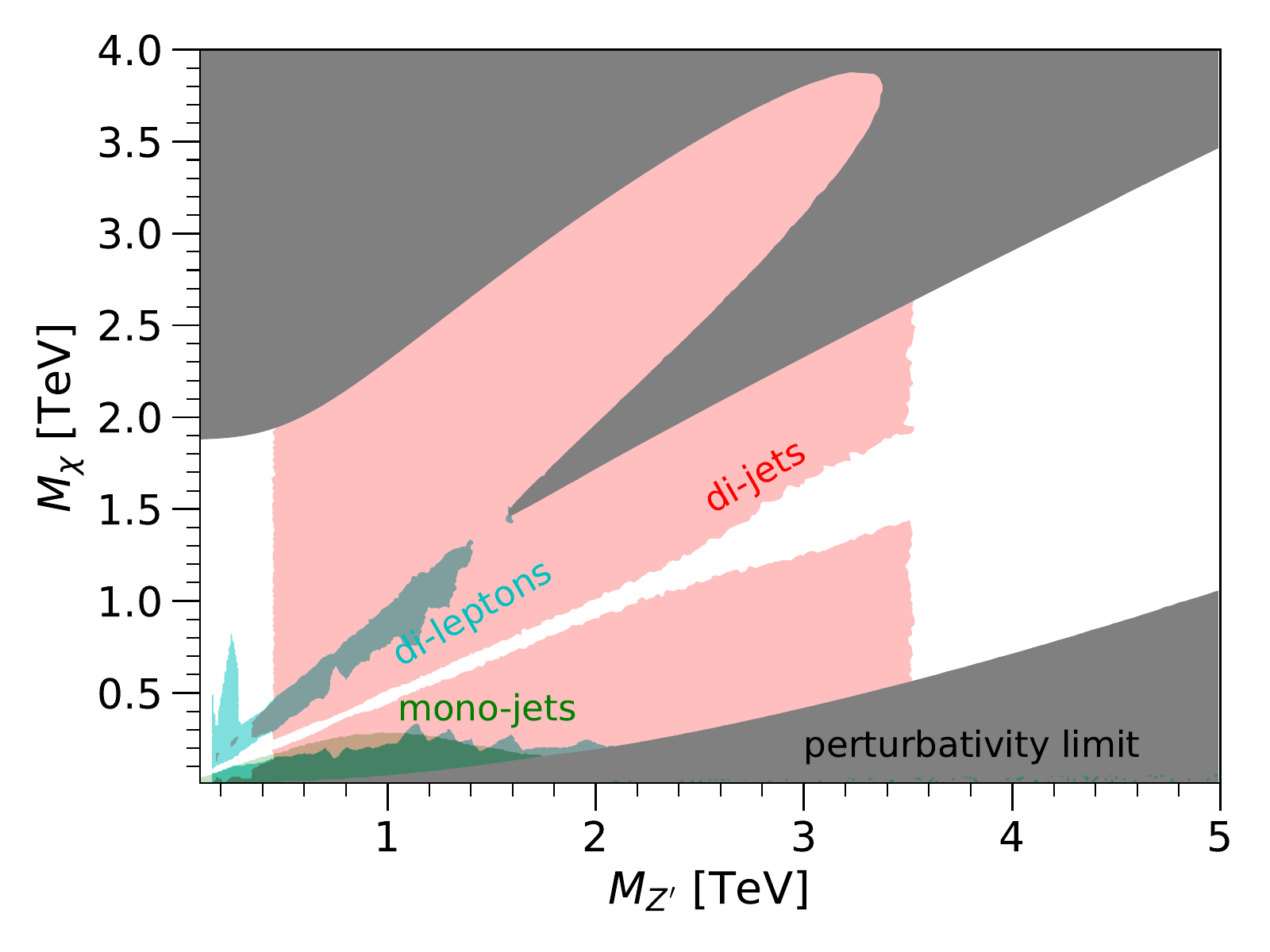}
\includegraphics[width=0.49\linewidth]{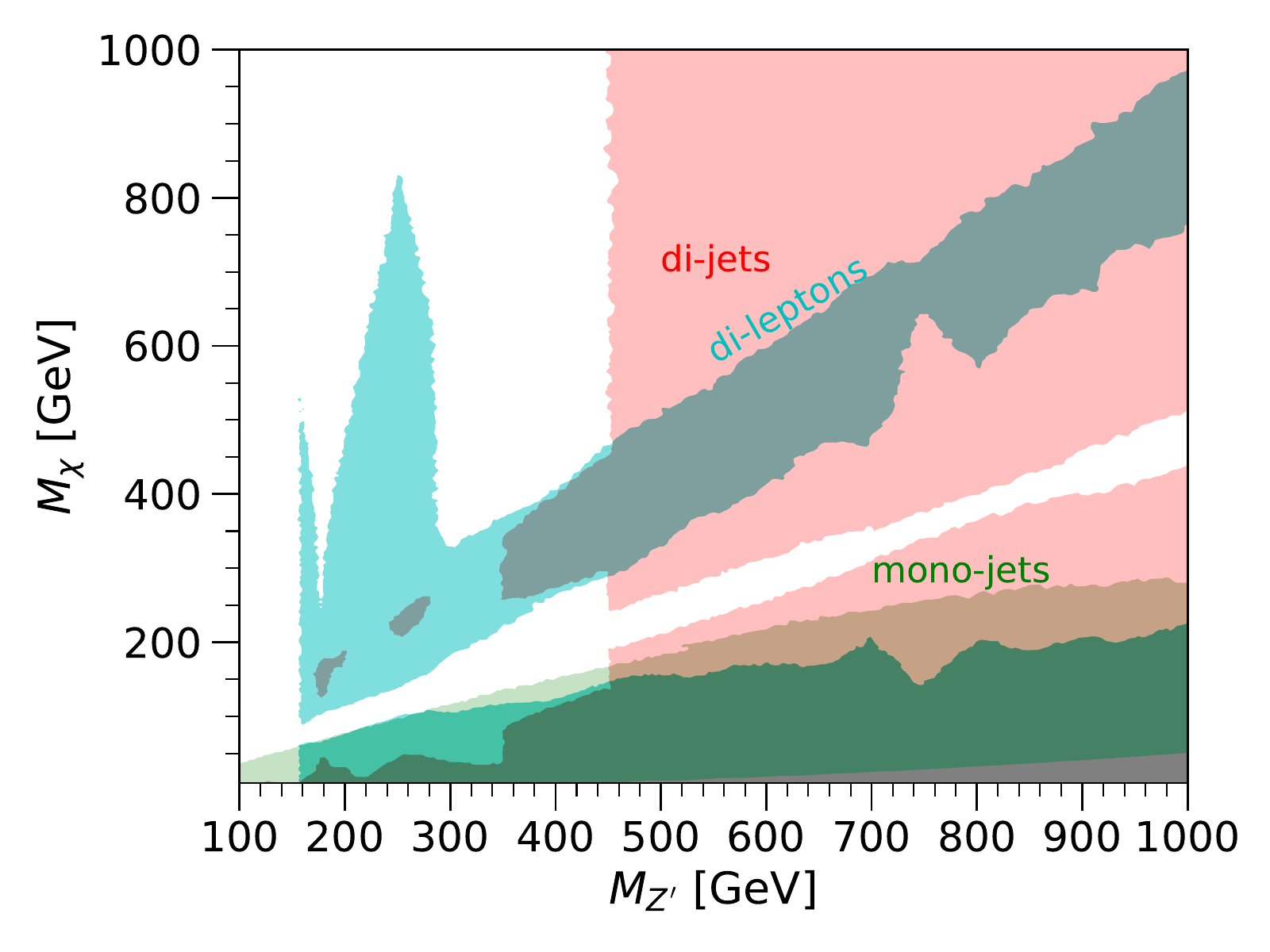}\\
\includegraphics[width=0.49\linewidth]{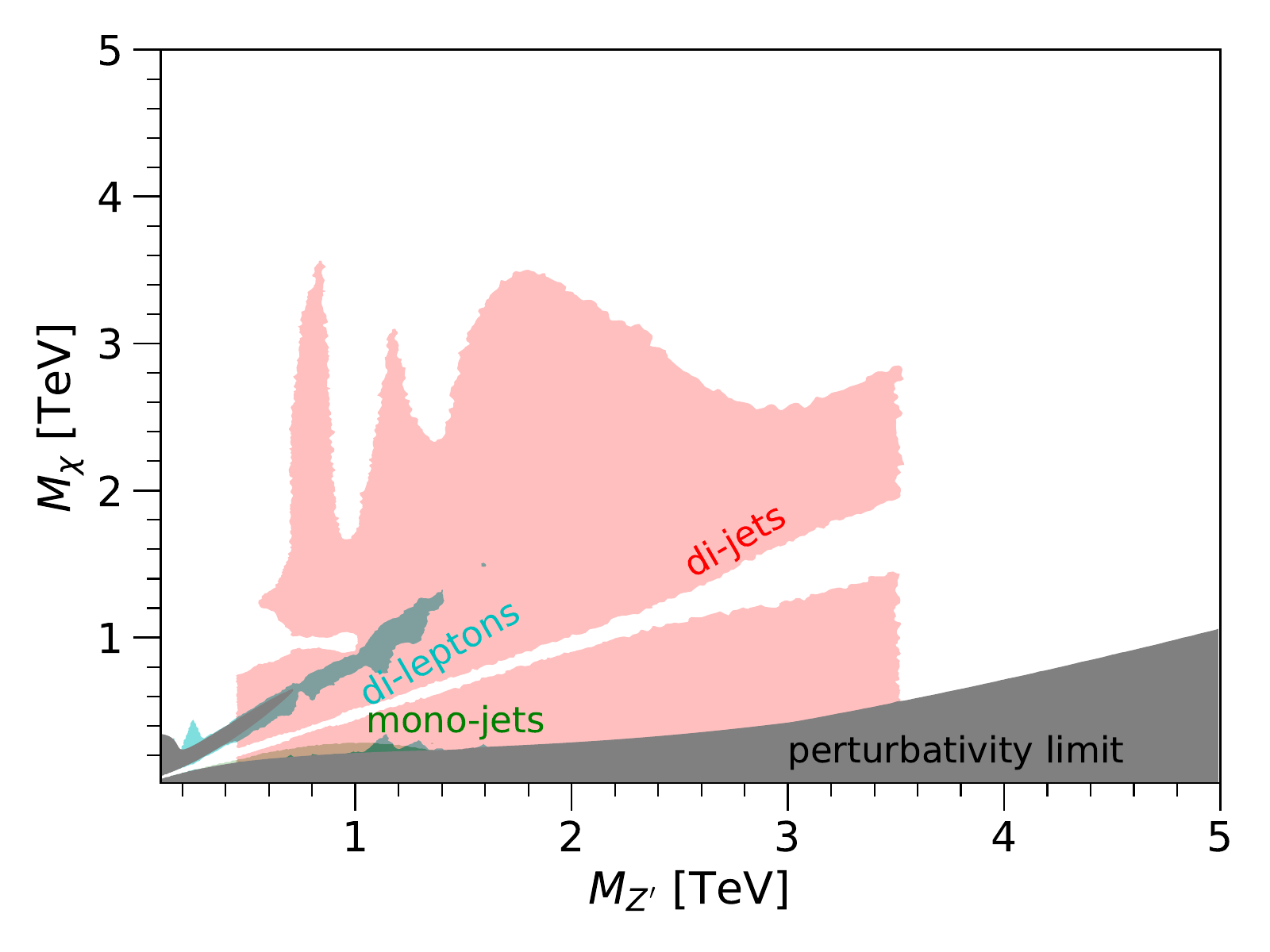}
\includegraphics[width=0.49\linewidth]{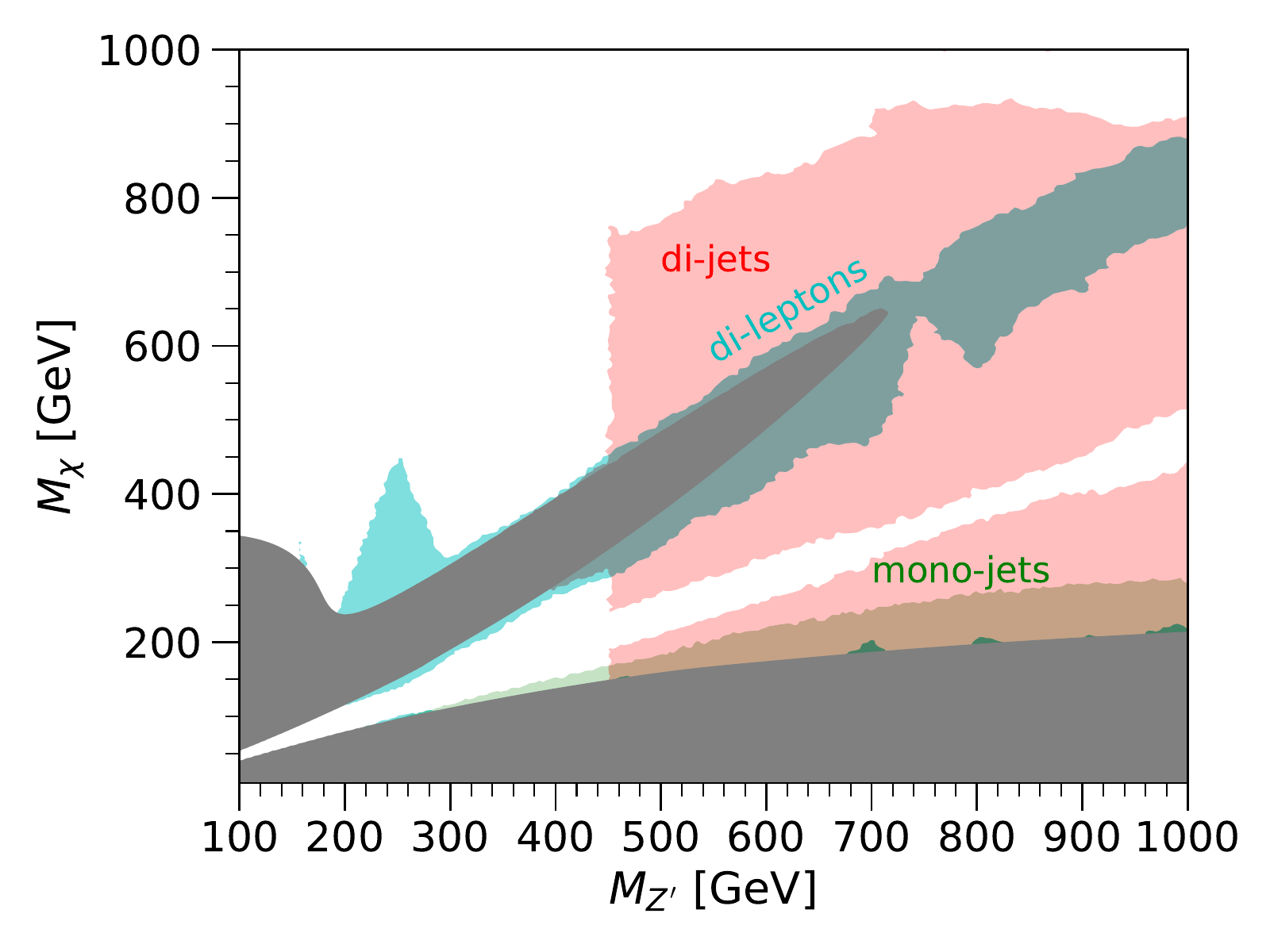}
\caption{
The same as Fig. \ref{fig:Lambdalow} for $\Lambda' = 100 \ m_{Z'}$. The turquoise-shaded area is excluded by di-lepton resonance searches. 
%The $m_{Z'}- m_\chi$ plane forbidden by constraints from di-lepton resonance searches (turquoise), di-jets (pink) and mono-jets (green) and the parameter space where perturbative is violated (grey). The value of the $g_B$ coupling is adjusted to reproduce the right DM relic density for $\Lambda' = 100 m_{Z'}$. The upper panels show the case where $m_s = 15$ TeV, left-handed showing the full range of $m_{Z'}$ considered while right-handed we zoom in the region of Z' masses up 1 TeV. The lower panels show the same but for the $m_s = 2$ TeV case.
}
\label{fig:Lambdahigh}
\end{figure}

For the value of $\Lambda'$ considered (a rather low one), the kinetic mixing is not sizeable and does not lead to relevant constraints from EWPO and di-lepton production. The corresponding bounds on the plane are close to the perturbativity one, and always weaker than other phenomenological constraints. For $m_Z' \simlt 500\ {\rm GeV}$ the most important bounds come from mono-jet production (green area). Still there is a lot of viable parameter space in this regime of relatively light $Z'$.
%{\red SC: Do we understand why there are no Z' limits below 500 GeV ? Is it not ruled out from Tevatron di-jet searches etc. RR: We have checked that Tevatron constraints from di-jet searches are weaker than ATLAS/CMS ones from mono-X searches for Z' of a few hundred GeVs. In any case we have added a line in the text mentioning this.}
Fig. \ref{fig:Lambdahigh} shows the constraints when the UV scale is large, $\Lambda'=100\ m_{Z'}$. Bounds from di-jets and mono-jets remain as before, since they are essentially independent of the kinetic mixing. However, bounds from di-leptons become now important in the region of light $Z'$, excluding new areas in that regime. In contrast, EWPO bounds remain unimportant.  Still, there remain large viable regions for $m_Z' \simlt 500\ {\rm GeV}$, especially for a not very heavy scalar (last two panels).
%
%%%%%%%%%%%%%%%%%%%%%%%
\section{Conclusions}
\label{sec:conclusions}
%%%%%%%%%%%%%%%%%%%%%%

The possibility that the DM particle interacts with the SM fields via a $Z'-$boson (`$Z'-$portal') remains one of the most attractive WIMP scenarios, both from the theoretical and the phenomenological points of view. However WIMP models are under increasing pressure, due, specially, to direct detection (DD) experiments. In the case of generic $Z'-$models, another critical constraint comes from from di-lepton production at the LHC. These constraints are highly alleviated if the coupling of $Z'$ with DM is of the axial type, and if $Z'$ has leptophobic couplings to the SM particles, respectively. Such conditions have been often considered in the context of simplified DM models, which, however, do not take into account the restrictions coming from anomaly-cancellation.

Following the point of view of ref.\cite{Ellis:2017tkh}, we have considered in this paper generic, anomaly-free, leptophobic models, later particularized to the case where the $Z'$ boson is axially coupled to (fermionic) DM. Leptophobia implies that the extra $U(1)$ factor is exactly as baryon number in the SM sector, provided it is flavour-blind (which is extremely desirable from FCNC constraints). Then, there are about four hundred models (with ${\cal O}(1)$ charges and minimal dark sector), from which only four present axial couplings in the DM sector. These four cases are in fact very similar. The dark sector consists of the DM fermion, $\chi$, plus a SU(2) doublet and a singlet, both with non-vanishing hypercharges. In addition there is the scalar, $S$, responsible for the breaking of the extra $U(1)$ and giving mass to the associated gauge boson ($\sim Z'$). The extra stuff in the dark sector can produce non-trivial phenomenology, both for DM annihilation at the early universe and collider processes. However, we have focused in the simplest case where the extra dark fermions are heavy enough to be integrated out, leaving a theory with the DM particle, $\chi$ and the $Z'$ boson (and possibly the $s$ scalar); with three parameters: the gauge coupling, $g_B$, and the two masses $m_{Z'}$ and $m_\chi$ (plus $m_s$ if the scalar is not too heavy).

The resulting scenario is well protected from DD bounds, as the vectorial (axial) coupling of $Z'$ to the SM (DM) sector leads to spin-dependent DD interactions, which are velocity-suppressed as well. The latter is also true for indirect detection processes. These are good news for the viability of the model. For each choice of $\{m_{Z'}, m_\chi, m_s\}$ there is a unique value of $g_B$ leading to the correct relic density, $\Omega_{\rm DM}$, normally in the perturbative regime. The main difference of this anomaly-free scenario with the previous generic simplified models is that the vectorial type of coupling of $Z'$ to quarks is mandatory and that the ratio of the SM and DM couplings of $Z'$ is fixed by the anomaly-cancellation condition, namely $g_q/g_{\rm DM}=2/9$. 

We have analyzed the main experimental constraints on the model. Two of them, di-lepton production at the LHC and contribution to EWPO (particularly, $S$ and $T$ parameters), come from the kinetic mixing between the $U(1)_Y$ and $U(1)_B$ gauge-bosons. Even if such mixing is initially vanishing (at some UV scale), it arises radiatively from loop-diagrams involving quarks. Di-lepton and EWPO bounds are specially significant in the region of light $Z'$. In addition, we have included bounds from mono-jet and di-jet production at the LHC. While the former are also specially relevant at low $m_{Z'}$, the latter is dominant in the $500\ {\rm GeV}\simlt m_Z' \simlt 3000\ {\rm GeV}$ region, which becomes essentially excluded, except around the resonances, $2m_\chi\sim m_{Z'}, m_s$. Still, there remain large viable regions for $m_Z' \simlt 500\ {\rm GeV}$.

The possibility to test a scenario of this kind necessarily involves collider experiments. E.g. from Figs. 5, 6, it is clear that in the next years the LHC is going to explore regions of the parameter space which are now allowed, hopefully giving a positive signal of a model of this kind. Notice also that a future measurement of the $g_q/g_{\rm DM}$ ratio would be a dramatic test of the scenario. In addition, one can consider a more generic scenario os this type, where the extra fermions in the dark sector are not that heavy, so that they change the DM phenomenology (e.g. through co-annihilation processes), as well as the LHC one, since these particles are non-trivial representations of the SM gauge group and can be produced in the LHC collisions. Work along these lines is already in progress.

When this work was completed there has appeared a paper by Ellis et. al. \cite{Ellis:2018xal} examining two leptophobic and two axial (DM) $Z'$ models in a similar spirit. The main difference with our case is that the scenario analyzed here is simultaneously leptophobic and axial.

\acknowledgments
This work has been partially supported by MINECO, Spain, under contracts FPA2014-57816-P, FPA2016-78022-P  and Centro de excelencia Severo Ochoa Program under grants SEV-2014-0398 and SEV-2016-0597; by the European Union projects H2020-MSCA-RISE-2015-690575-InvisiblesPlus and H2020-MSCA-ITN-2015/674896-ELUSIVES; and by Generalitat Valenciana grant PROMETEOII\-2014/050. The work of J.Q. is supported through the Spanish FPI grant SVP-2014-068899. 
J.A.C. thanks Jes\'us M. Moreno and J. Antonio Aguilar-Saavedra for useful discussions.
J.Q. thanks Professor Alejandro Ibarra and Physik-Department T30d from the Technische Universit$\ddot {\textrm a}$t M$\ddot {\textrm u}$nchen for the hospitality during  a big part of the execution of this work. J.Q. also thanks Professor Alejandro Ibarra for interesting discussions. 
J.Q. thanks Claudia Garc\'ia Garc\'ia for her invaluable help with computational resources.

\newpage

\appendix
\section{Apendix: Anomaly-free completions of $U(1)_B$}

As discussed in section 2 any consistent leptophobic, flavour-blind, $U(1)_{Y'}$ group must be equivalent to baryonic number, $U(1)_B$, in the SM sector. Furthermore, anomaly-cancellation requires the presence of extra particles. Then, assuming that the DM particle, $\chi$, is a fermion with vanishing hypercharge, the minimal content of the dark sector contains an additional doublet, $\psi$ and an additional singlet, $\eta$:
\be
{\rm  minimal\ dark\ sector:} \ \ \ \{\chi_{L,R},\  \psi_{L,R},\  \eta_{L,R}\} ,
\label{minDS2}
\ee

In this appendix we fully classify the possible assignments of $Y, Y'$ to these fields, consistent with anomaly-cancellation, paying special attention to the axial cases \footnote{ There is relevant previous literature in this subject \cite{Pais:1973mi,Rajpoot:1987yg,Foot:1989ts,Carone:1995pu,Georgi:1996ei,Dulaney:2010dj,FileviezPerez:2010gw,FileviezPerez:2011pt,Arnold:2013qja,Duerr:2013lka,Perez:2013tea,Batell:2014yra,Duerr:2014wra,Perez:2015rza,Duerr:2015vna,Ellis:2018xal}. Here we supplement previous classifications with cases that were not considered and make explicit the form of all solutions, with special focus on the axial case.}.
Notice that the requirement of non-fractional electric charges implies $Y_\psi=m+1/2,Y_\eta=n$, with $m,n$ integers, a condition that we will assume in what follows.

A useful observation is that the anomaly-cancelation conditions, listed in the equations (\ref{SU2Yp}-\ref{Yp3}), are invariant under the three independent transformations:
\be
Y_{\psi,\eta}\rightarrow - Y_{\psi,\eta}
\label{sym1}
\ee
\be
Y'_{(\psi,\eta)_L}\leftrightarrow - Y'_{(\psi,\eta)_R}
\label{sym2}
\ee
\be
Y'_{\chi_L}\leftrightarrow - Y'_{\chi_R}
\label{sym3}
\ee
Hence, in general the solutions to the anomaly-cancellation conditions come in sets of 8 possibilities related by these transformations. 

\subsection{Classification of solutions}

In this subsection will derive the possible values of the extra hypercharges ($Y'$) of the fields in the dark sector (\ref{minDS2}) for any choice of $Y_{\psi}, Y_{\eta}$.

From Eqs.(\ref{SU2Yp}-\ref{Y2Yp}) we can solve ${Y'_{\psi_L}, Y'_{\eta_L}, Y'_{\chi_L}}$ in terms of the other charges:
\bea
Y'_{\psi_L}&=&Y'_{\psi_R}-3
\nonumber\\
Y'_{\eta_L}&=&Y'_{\eta_R}+\frac{3}{2Y_\eta^2}(1+4Y_\psi^2)
\nonumber\\
Y'_{\chi_L}&=&Y'_{\chi_R}-\frac{3}{2Y_\eta^2}(1+4Y_\psi^2)+6
\label{solve1}
\eea
%
%At this stage it is already clear that there are two choices of $Y_{\psi}, Y_{\eta}$ that lead to a very symmetric form of the solutions. Namely, for $\left\{Y_{\psi}, Y_{\eta}\right\}=\left\{\frac{1}{2}, 1\right\}, \left\{\frac{7}{2}, 5\right\}$ (and the solutions obtained from (\ref{sym1})), it happens that $Y'_{(\psi_R, \eta_L, \chi_L)}=Y'_{(\psi_L, \eta_R, \chi_R)}+3$. We will see shortly that these special choices have other interesting features.
%
The value of $Y'_{\eta_R}$ can be derived from Eq.(\ref{Yp2Y}), which, thanks to Eqs.(\ref{solve1}) becomes linear in $Y'_{\psi_R}$:
\bea
Y'_{\eta_R}=\frac{2Y_\eta(-3+2Y'_{\psi_R})}{1+4Y_\psi^2} - \frac{3(1+4Y_\psi^2)}{4Y_\eta^ 2}
\label{solve2}
\eea
So far we have expressed ${Y'_{\psi_L}, Y'_{\eta_L}, Y'_{\chi_L}, Y'_{\eta_R}}$ in terms of 
${Y_{\psi}, Y_{\eta}, Y'_{\psi_R},Y'_{\chi_R}}$. Now, for a given choice of $Y_{\psi}, Y_{\eta}$, the values of $Y'_{\psi_R},Y'_{\chi_R}$ are related by the only remaining anomaly-cancelation condition, namely Eq.(\ref{Yp3}), which, thanks to Eqs.(\ref{solve1}) becomes quadratic in the unknowns:
\bea
\frac{1}{32Y_\eta^ 6(1+4Y_\psi^2)}
&&\left\{
9(-16Y_\eta^4(6+Y'_{\chi_R})^2)(1+4Y_\psi^2)^2
+24Y_\eta^2(6+Y'_{\chi_R}))(1+4Y_\psi^2)^3 -9(1+4Y_\psi^2)^4
\right.
\nonumber\\
&&
\hspace{-1cm}
\left.
-64Y_\eta^6( 9-(-3+Y'_{\psi_R})Y'_{\psi_R}  +45 Y_\psi^2 +
Y'_{\chi_R}  (6+Y'_{\chi_R}))(1+4Y_\psi^2)  )
\right\}
=0
\label{solve3}
\eea
Consequently, one would expect that for any choice of $Y_{\psi}, Y_{\eta}$ there is a continuum of solutions. Still one has to require that these solutions are real. Let us examine closely this issue. Solving $Y'_{\chi_R}$ in Eq.(\ref{solve3}) gives
\bea
Y'_{\chi_R}=3\left(-1+\frac{1+4Y_\psi^2}{4Y_\eta^2}\right)\pm 
\frac{\sqrt{D}}{4Y_\eta^2 (1+4Y_\psi^2)(-1+4Y_\eta^2-4Y_\psi^2)}
\label{solve4}
\eea
with
\bea
D=&-&\frac{1}{Y_\eta^6}(-1+4Y_\eta^2-4Y_\psi^2)(1+4Y_\psi^2)
\nonumber\\
&\times&
\left[
-16Y_\eta^4(  (-3+Y'_{\psi_R})Y'_{\psi_R} -9  Y_{\psi}) + 9 (1+4Y_\psi^2)^3
-36(Y_\eta+4Y_\eta Y_\psi^2)^2
\right]
\label{D}
\eea
Obviously, real solutions correspond to $D\geq 0$. Let us note that the extremal point of the quadratic expression (\ref{D}) always lies at $Y'_{\psi_R}=3/2$ (this is a consequence of the symmetry (\ref{sym2}) and the first equation of (\ref{solve1})). At this extremal point $D$ reads
\bea
D^{\rm extr}=-\frac{9}{Y_\eta^6}(-1+4Y_\eta^2-4Y_\psi^2)(1+4Y_\psi^2) (1-2Y_\eta^2+4Y_\psi^2)^2
\label{Dextr}
\eea
On the other hand, the coefficient of $(Y'_{\psi_R})^2$ in (\ref{D}) reads
\bea
\frac{16}{Y_\eta^2}(-1+4Y_\eta^2-4Y_\psi^2)(1+4Y_\psi^2) 
\label{coeff}
\eea
Since expressions (\ref{Dextr}) and (\ref{coeff}) have opposite signs, it turns out that for any choice of $Y_{\psi}, Y_{\eta}$ there is indeed a continuum of values of $Y'_{\psi_R}$ that lead to real solutions:

\vspace{0.3cm}

If  $-1+4Y_\eta^2-4Y_\psi^2>0$, \hspace{1cm} $Y'_{\psi_R}\leq Y'^{\ (1)}_{\psi_R}$ ~  \& ~ $Y'_{\psi_R}\geq Y'^{\ (2)}_{\psi_R}$

\vspace{0.3cm}

If  $-1+4Y_\eta^2-4Y_\psi^2<0$, \hspace{1cm} $Y'^{\ (1)}_{\psi_R}\leq Y'_{\psi_R}\leq Y'^{\ (2)}_{\psi_R}$

\vspace{0.3cm}
\noindent
where
\bea
Y'^{\ (1,2)}_{\psi_R}=\frac{3}{2}\mp\frac{3|1-2Y_\eta^2+4Y_\psi^2|}{4Y_\eta^2}\sqrt{
(1+4Y_\psi^2)
}
\label{Y12}
\eea
Then, for each allowed value of $Y'_{\psi_R}$, the corresponding value of $Y'_{\chi_R}$ is given by Eq.(\ref{solve4}). %It should be noticed that in general the solutions are not rational in both. 

\subsection{ Special Choices of $Y_{\psi}, Y_{\eta}$}

There are four special choices of $Y_{\psi}, Y_{\eta}$ that lead to a substantial simplification of the solutions and, besides, allow for generic rational solutions.
Namely, for 
\bea
\left\{\pm Y_{\psi}, \pm Y_{\eta}\right\}=\left\{\frac{1}{2}, 1\right\}, \left\{\frac{7}{2}, 5\right\}\ ,
\label{special}
\eea
Eqs.(\ref{solve1}) become
\bea
Y'_{\psi_L}&=&Y'_{\psi_R}-3
\nonumber\\
Y'_{\eta_L}&=&Y'_{\eta_R}+ 3
\nonumber\\
Y'_{\chi_L}&=&Y'_{\chi_R}+3
\label{solve1E}
\eea
The value of $Y'_{\eta_R}$ becomes
\bea
&&Y'_{\eta_R}=\frac{1}{8}(-24+8Y'_{\psi_R}),\ \ \ \ {\rm for}\ \ \left\{\pm Y_{\psi}, \pm Y_{\eta}\right\}=\left\{\frac{1}{2}, 1\right\}
\nonumber\\
&&Y'_{\eta_R}=\frac{1}{5}(-18+7Y'_{\psi_R}),\ \ \ \ {\rm for}\ \ \left\{\pm Y_{\psi}, \pm Y_{\eta}\right\}=\left\{\frac{7}{2},
5\right\}
\label{solve2E}
\eea
The value of $Y'_{\chi_R}$, Eq.(\ref{solve4}), gets also drastically simplified:
\bea
&&Y'_{\chi_R}=-3+Y'_{\psi_R}, \ \ -Y'_{\psi_R}\ \ ,\ \ \ \ {\rm for}\ \ \left\{\pm Y_{\psi}, \pm Y_{\eta}\right\}=\left\{\frac{1}{2}, 1\right\}
\nonumber\\
&&Y'_{\chi_R}=\frac{1}{5}(-6-Y'_{\psi_R}),\ \ \frac{1}{5}(-9+Y'_{\psi_R})\ \ ,\ \ \ \ {\rm for}\ \ \left\{\pm Y_{\psi}, \pm Y_{\eta}\right\}=\left\{\frac{7}{2},
5\right\}
\label{solve4E}
\eea
Not that, in each case, the two solutions for $Y'_{\chi_R}$ are related by the symmetry (\ref{sym3}) and Eq.(\ref{solve1E}).

A crucial consequence of the previous equations is that, in the special cases (\ref{special}), for any rational choice of $Y'_{\psi_R}$, the rest of the $Y'-$charges become rational as well. This cannot be guaranteed for any other choice of $Y_{\psi}, Y_{\eta}$. As a matter of fact, in general it does not hold, except by accident. In Table \ref{ExceptionalSolutions} we list accidental rational possibilities, which do not belong to the special choices (\ref{special}).

\begin{table}[htbp]
\begin{center}
\begin{tabular}{| c | c | c | c | c | c | c | c | c | }
\hline
$Y_{\psi}$ & $Y_{\eta}$ & $Y^\prime_{\psi_L}$ & $Y^\prime_{\psi_R}$ & $Y^\prime_{\eta_L}$ & $Y^\prime_{\eta_R}$ & $Y^\prime_{\chi_L}$ & $Y^\prime_{\chi_R}$ \\
\hline 
3/2 & 1 & -9 & 0 & -9 & -6 & 3 & -12 \\
\hline
3/2 & 1 & 3/8 & 75/8 & 3/8 & 27/8 & 69/8 & -51/8 \\
\hline
3/2 & 2 & 3/8 & -15/8 & 3/8 & 27/8 & 33/8 & 3/8 \\
\hline
3/2 & 3 & 5/3 & -8/3 & -4 & -1 & -11/3 & -16/3 \\
\hline
\end{tabular}
\caption{Accidental rational solutions to the anomaly equations. From each case, there are seven additional solutions, which can be obtained by using the transformations (\ref{sym1}-\ref{sym3}).}
\label{ExceptionalSolutions}
\end{center}
\end{table}

Some of the previous features come from the fact the special choices (\ref{special}) are the only ones for which $1-2Y_\eta^2+4Y_\psi^2=0$. This also implies that $D^{\rm extr}=0$ in Eq.(\ref{Dextr}). Since, on the other hand, $-1+4Y_\eta^2-4Y_\psi^2>0$, it turns out that all values of $Y'_{\psi_R}$ are allowed, in particular all rationals.

\subsection{Axial coupling of the dark matter}

In this subsection we particularise to the case where the coupling of the extra gauge boson to the dark matter is axial, i.e. 
\bea
Y'_{\chi_L}=-Y'_{\chi_R}.
\label{axialDM}
\eea
Let us start by noting that the two generic solutions of $Y'_{\chi_R}$ given in Eq.(\ref{solve4}) are related by the 
symmetry transformation (\ref{sym3}). Therefore, the axial case (\ref{axialDM}) occurs when 
 the two solutions coincide, i.e. when $D=0$. This happens precisely for $Y'_{\psi_R}=Y'^{\ (1)}_{\psi_R}, Y'^{\ (2)}_{\psi_R}$, given in Eq.(\ref{Y12}). 
 
Consequently, for any choice of $Y_{\psi}, Y_{\eta}$, there are two solutions of axial DM, with $Y'_{\psi_R}$ given by Eq.(\ref{Y12}); $Y'_{\chi_R}$, given by Eq.(\ref{solve4}), which in this case simplifies to
\bea
Y'_{\chi_R}=3\left(-1+\frac{1+4Y_\psi^2}{4Y_\eta^2}\right)
\label{solve4A}
\eea
 and the remaining charges given by Eqs.(\ref{solve1}, \ref{solve2}). 
 
Notice that the two values $Y'^{\ (1)}_{\psi_R}, Y'^{\ (2)}_{\psi_R}$ are symmetrical with respect to $Y'_{\psi_R}=3/2$ (as implied by the symmetry (\ref{sym1}) and Eq.(\ref{solve1})). This means that the solutions are {\em not} axial for the other dark fields, $\psi$ and $\eta$, {\em except} in the special cases (\ref{special}), where $Y'^{\ (1)}_{\psi_R}=Y'^{\ (2)}_{\psi_R}=3/2$. For each of these special cases there  is a unique axial solution, which, in addition, is  axial in all the dark fields as well. These are the ones given in Eq.(\ref{leptophobax}) of the section 2. Note also that these are the only axial solutions whose charges are rational.

 \newpage
 
\bibliographystyle{model1-num-names}
\bibliography{main.bib}

\end{document}